\documentclass[twocolumn,iop,numberedappendix,appendixfloats]{openjournal}
\usepackage[utf8]{inputenc}
\usepackage{amsmath,amssymb,amstext}
\usepackage{bm}
\usepackage{graphicx}
\usepackage{booktabs} 
\usepackage{acronym} 

\graphicspath{ {./images/} }
\usepackage{natbib}
\usepackage{subcaption}
\usepackage[colorlinks,allcolors=blue]{hyperref}
\usepackage{dsfont}
\usepackage{comment}
\usepackage{float}
\usepackage{xcolor}
\usepackage{listings}
\usepackage{fontawesome}
\usepackage{float}

\usepackage{tikz}
\usetikzlibrary{shapes,arrows,shadows,fit}
\usetikzlibrary{positioning}
\usetikzlibrary{bayesnet}

\definecolor{maroon}{cmyk}{0, 0.87, 0.68, 0.32}
\definecolor{halfgray}{gray}{0.55}
\definecolor{ipython_frame}{RGB}{207, 207, 207}
\definecolor{ipython_bg}{RGB}{247, 247, 247}
\definecolor{ipython_red}{RGB}{186, 33, 33}
\definecolor{ipython_green}{RGB}{0, 128, 0}
\definecolor{ipython_cyan}{RGB}{64, 128, 128}
\definecolor{ipython_purple}{RGB}{170, 34, 255}
\definecolor{another_red}{RGB}{235, 64, 52}

\usepackage{mdframed}

\lstdefinelanguage{iPython}{
    morekeywords={access,and,break,class,continue,def,del,elif,else,except,exec,finally,for,from,global,if,import,in,is,lambda,not,or,pass,print,raise,return,try,while},%
    morekeywords=[2]{abs,all,any,basestring,bin,bool,bytearray,callable,chr,classmethod,cmp,compile,complex,delattr,dict,dir,divmod,enumerate,eval,execfile,file,filter,float,format,frozenset,getattr,globals,hasattr,hash,help,hex,id,input,int,isinstance,issubclass,iter,len,list,locals,long,map,max,memoryview,min,next,object,oct,open,ord,pow,property,range,raw_input,reduce,reload,repr,reversed,round,set,setattr,slice,sorted,staticmethod,str,sum,super,tuple,type,unichr,unicode,vars,xrange,zip,apply,buffer,coerce,intern, function, @model, Uniform, Normal, MvNormal, theory_planck},%
    sensitive=true,%
    morecomment=[l]\#,%
    morestring=[b]',%
    morestring=[b]",%
    morestring=[s]{'''}{'''},
    morestring=[s]{"""}{"""},
    morestring=[s]{r'}{'},
    morestring=[s]{r"}{"},%
    morestring=[s]{r'''}{'''},%
    morestring=[s]{r"""}{"""},%
    morestring=[s]{u'}{'},
    morestring=[s]{u"}{"},%
    morestring=[s]{u'''}{'''},%
    morestring=[s]{u"""}{"""},%
    literate=
    {á}{{\'a}}1 {é}{{\'e}}1 {í}{{\'i}}1 {ó}{{\'o}}1 {ú}{{\'u}}1
    {Á}{{\'A}}1 {É}{{\'E}}1 {Í}{{\'I}}1 {Ó}{{\'O}}1 {Ú}{{\'U}}1
    {à}{{\`a}}1 {è}{{\`e}}1 {ì}{{\`i}}1 {ò}{{\`o}}1 {ù}{{\`u}}1
    {À}{{\`A}}1 {È}{{\'E}}1 {Ì}{{\`I}}1 {Ò}{{\`O}}1 {Ù}{{\`U}}1
    {ä}{{\"a}}1 {ë}{{\"e}}1 {ï}{{\"i}}1 {ö}{{\"o}}1 {ü}{{\"u}}1
    {Ä}{{\"A}}1 {Ë}{{\"E}}1 {Ï}{{\"I}}1 {Ö}{{\"O}}1 {Ü}{{\"U}}1
    {â}{{\^a}}1 {ê}{{\^e}}1 {î}{{\^i}}1 {ô}{{\^o}}1 {û}{{\^u}}1
    {Â}{{\^A}}1 {Ê}{{\^E}}1 {Î}{{\^I}}1 {Ô}{{\^O}}1 {Û}{{\^U}}1
    {œ}{{\oe}}1 {Œ}{{\OE}}1 {æ}{{\ae}}1 {Æ}{{\AE}}1 {ß}{{\ss}}1
    {ç}{{\c c}}1 {Ç}{{\c C}}1 {ø}{{\o}}1 {å}{{\r a}}1 {Å}{{\r A}}1
    {€}{{\EUR}}1 {£}{{\pounds}}1
    {^}{{{\color{ipython_purple}\^{}}}}1
    {=}{{{\color{ipython_purple}=}}}1
    {+}{{{\color{ipython_purple}+}}}1
    {-}{{{\color{ipython_purple}-}}}1
    {*}{{{\color{ipython_purple}$^\ast$}}}1
    {/}{{{\color{ipython_purple}/}}}1
    {+=}{{{+=}}}1
    {-=}{{{-=}}}1
    {*=}{{{$^\ast$=}}}1
    {/=}{{{/=}}}1,
    literate=
    *{-}{{{\color{ipython_purple}-}}}1
     {?}{{{\color{ipython_purple}?}}}1,
    identifierstyle=\color{black}\ttfamily,
    commentstyle=\color{ipython_cyan}\ttfamily,
    stringstyle=\color{ipython_red}\ttfamily,
    keepspaces=true,
    showspaces=false,
    showstringspaces=false,
    rulecolor=\color{ipython_frame},
    frameround={t}{t}{t}{t},
    numbers=none,
    numberstyle=\tiny\color{halfgray},
    backgroundcolor=\color{ipython_bg},
    basicstyle=\ttfamily\footnotesize,
    columns=fullflexible,
    keywordstyle=\color{ipython_green}\ttfamily,
}

\newcommand{\Planck}{Planck}
\newcommand{\flinch}{\texttt{Flinch.jl}}
\newcommand{\capse}{\texttt{Capse.jl}}

\newcommand{\zygote}{\texttt{Zygote.jl}}
\newcommand{\julia}{\texttt{Julia}}
\newcommand{\healpix}{\texttt{HEALPix}}
\newcommand{\healpixmpi}{\texttt{HealpixMPI.jl}}
\newcommand{\Cls}{\(C_\ell\)'s}
\newcommand{\alm}{\(a_{\ell m}\)}
\newcommand{\almtomap}{\texttt{alm2map}}
\newcommand{\almtomapPB}{\texttt{alm2map\_pullback}}
\newcommand{\adjalmtomap}{\texttt{adjoint\_alm2map}}
\newcommand{\almanac}{\texttt{Almanac}}
\newcommand{\namaster}{\texttt{NaMaster}}

\date{\today}

\begin{document}
\journalinfo{The Open Journal of Astrophysics}
\submitted{submitted XXX; accepted YYY}

\shorttitle{\flinch{}: Towards optimal inference of angular maps}
\shortauthors{}
\title{\flinch{}: A Differentiable Framework for Field-Level Inference of Cosmological parameters from curved sky data}
\author{Andrea Crespi$^{\star1,2,3}$}
\author{Marco Bonici$^{1,2}$}
\author{Arthur Loureiro $^{4,5}$}
\author{Jaime Ruiz-Zapatero $^{6,7}$}
\author{Ivan Sladoljev $^{4,8}$}
\author{Zack Li $^{9,10}$}
\author{Adrian Bayer $^{11, 12}$}
\author{Marius Millea$^{10, 13}$}
\author{Uro\v{s} Seljak $^{9,10,14}$}
\affiliation{$^1$ Waterloo Centre for Astrophysics, University of Waterloo, Waterloo, ON N2L 3G1, Canada}
\affiliation{$^2$ Department of Physics and Astronomy, University of Waterloo, Waterloo, ON N2L 3G1, Canada}
\affiliation{$^3$ Dipartimento di Fisica Aldo Pontremoli, UNIMI}
\affiliation{$^4$ Oskar Klein Centre for Cosmoparticle Physics, Department of Physics, Stockholm University, Stockholm, SE-106 91, Sweden}
\affiliation{$^5$ Astrophysics Group, Blackett Laboratory, Imperial College London, London SW7 2AZ, UK}
\affiliation{$^6$ Department of Physics and Astronomy, University College London, Gower Street, London WC1E 6BT, UK}
\affiliation{$^7$ Advanced Research Computing Centre, University College London, 90 High Holborn, London WC1V 6LJ, UK}
\affiliation{$^8$ Department of Physics, Royal Holloway, University of London, Egham Hill, Egham, UK}
\affiliation{$^9$ Department of Physics, University of California, Berkeley, CA 94720, USA}
\affiliation{$^{10}$ Berkeley Center for Cosmological Physics, UC Berkeley, CA 94720, USA}
\affiliation{$^{11}$ Center for Computational Astrophysics, Flatiron Institute, 162 5th Avenue, New York, NY 10010, USA}
\affiliation{$^{12}$ Department of Astrophysical Sciences, Princeton University, Peyton Hall, Princeton, NJ 08544, USA}
\affiliation{$^{13}$ Department of Physics, University of California, Davis, CA 95616, USA}
\affiliation{$^{14}$ Lawrence Berkeley National Laboratory, One Cyclotron Road, Berkeley, CA 94720, USA}
\thanks{$^\star$ E-mail: \nolinkurl{a2crespi@uwaterloo.ca} }

\begin{abstract}
We present \flinch{}, a fully differentiable and high-performance framework for field-level inference on angular maps, developed to improve the flexibility and scalability of current methodologies. \flinch{} is integrated with differentiable cosmology tools, allowing gradients to propagate from individual map pixels directly to the underlying cosmological parameters. This architecture allows cosmological inference to be carried out directly from the map itself, bypassing the need to specify a likelihood for intermediate summary statistics. Using simulated, masked CMB temperature maps, we validate our pipeline by reconstructing both maps and angular power spectra, and we perform cosmological parameter inference with competitive precision. In comparison with the standard pseudo-\Cls{} approach, \flinch{} delivers substantially tighter constraints, with error bars reduced by up to 40\%. Among the gradient-based samplers routinely employed in field-level analyses, we further show that MicroCanonical Langevin Monte Carlo provides orders-of-magnitude improvements in sampling efficiency over currently employed Hamiltonian MonteCarlo samplers, greatly reducing computational expense.
\end{abstract}

\maketitle
\section{Introduction}
\label{sec:introduction}

The current and coming decade will confront cosmology with an unprecedented avalanche of precision data, from the galaxy–redshift cartography of DESI~\citep{DESI:2025fxa}, Euclid~\citep{Euclid:2024yrr}, and the Rubin Observatory Legacy Survey of Space and Time (LSST)~\citep{Ivezic2019LSST:Products} to the polarized microwave skies of the Simons Observatory~\citep{SimonsObservatory:2018koc} and LiteBIRD~\citep{LiteBIRD:2024wix}. Extracting the full information content from these surveys is a formidable statistical and computational challenge: for decades, cosmologists have exploited the near-Gaussianity of cosmological fields by compressing data into two-point correlation statistics, either the real-space correlation function or its harmonic-space counterpart, and used these to infer the properties of the cosmos~\citep{Tegmark:1996qt, Tegmark:2001zv, Bond:1994aa, Gorski:1994ye, Hamilton:2005kz, Hamilton:2005ma, Alonso:2018jzx, Philcox:2020vbm, Philcox:2021ukg}. This paradigm has driven flagship analyses of CMB temperature and polarization maps, cosmic-shear catalogs, and galaxy-clustering samples, underpinning the precision constraints reported by experiments such as Planck~\citep{Planck:2018vyg}, KiDS~\citep{Wright:2025xka}, DES~\citep{DES:2021wwk}, BOSS~\citep{BOSS:2016wmc}, and DESI~\citep{DESI:2025fxa}, thereby sharpening our picture of the Universe.

Despite their impressive track record, two-point statistics are only the first rung of a much taller information ladder. In recent years, a concerted effort to capture the non-Gaussian information discarded by the power spectrum has spurred the development of a diverse suite of “beyond-2pt” techniques. These range from direct extensions to higher-order $N$-point functions such as the bispectrum~\citep{Philcox:2021kcw,DAmico:2022osl} to novel summaries based on wavelets, voids, and other features of the cosmic web~\citep{Hahn:2022wgo, Bayer:2021iyb, Euclid:2021xmh,Euclid:2022qtk,Euclid:2022hdx, Paillas:2023cpk, DES:2024jgw,Valogiannis:2024rvt, Sunseri:2025jem}, whose growing maturity is evidenced by community-wide validation efforts such as the Beyond-2pt mock data challenge~\citep{Beyond-2pt:2024mqz}.

While powerful, these methods still rely on some form of data compression. A natural endpoint of this trajectory is to eschew summary statistics altogether and perform inference directly on the pixelized data maps, a paradigm known as field-level inference (FLI). By operating on the full field, FLI can, in principle, capture information from all $N$-point correlation functions simultaneously \citep{2021MNRAS.506L..85L, SpurioMancini:2024qic}, while naturally accommodating complex observational effects such as survey masks, inhomogeneous noise, and instrumental beams. Under a correct generative model, this approach can be statistically optimal in the sense that no information is lost to compression; in practice, its advantages must be balanced against the computational challenges posed by high-dimensional and potentially non-Gaussian posteriors.

Within FLI, two complementary strategies have emerged. The first is a physics-informed forward-modeling approach, exemplified by \texttt{BORG}~\citep{2013MNRAS.432..894J}, \texttt{pmwd}~\citep{Li:2022qlf}, and \texttt{LEFTfield}~\citep{Schmidt:2020ovm}, which posits a dynamical model of gravitational structure formation (e.g., $N$-body or perturbative solvers) to evolve a set of initial conditions forward in time, fitting the resulting density field directly to observations. The second, more agnostic, strategy is a statistical hierarchical approach, typified by \almanac{}~\citep{Loureiro_2023, Sellentin_2023}, in which the latent cosmological fields and their power spectra are treated as stochastic variables, rather than being linked by a deterministic structure-formation map, an approach we refer to as \textit{statistical field-level inference}. Even for Gaussian random fields, there is merit in an \almanac{}-like approach: in this regime the commonly employed pseudo-$C_\ell$ estimator is optimal only for flat power spectra~\citep{1973ApJ...185..413P, Efstathiou:2003dj, Leistedt:2013gfa, Alonso:2018jzx}. This is particularly pertinent at the lowest multipoles, where the sampling distribution of the $C_\ell$ departs from Gaussianity (being closer to a Wishart form), rendering Gaussian-likelihood approximations inaccurate~\citep{Hamimeche:2008ai,carrongaussian,Oehl:2024gbm}. Such effects can especially degrade constraints on parameters like $f_\mathrm{NL}$, whose signal is concentrated on the largest scales~\citep{Dalal:2007cu, 2023MNRAS.520.5746A, Cagliari:2023mkq, DESI:2023duv, Chaussidon:2024qni, Cagliari:2025rqe, fabbianquaia}.

While several pipelines for both physics-informed and statistical FLI exist, a central obstacle remains: the sheer dimensionality of the problem. FLI posteriors routinely span millions of latent degrees of freedom, intertwined with hyperparameters and nuisance terms, yielding sharply curved, highly correlated, and often non-Gaussian target distributions. Scaling inference to this regime typically necessitates gradient-based samplers such as Hamiltonian Monte Carlo (HMC) and its variants, along with careful blocking, preconditioning, and mass-matrix adaptation to achieve acceptable mixing and wall-clock efficiency.

A second, compounding challenge is the evaluation of gradients themselves. Many existing FLI frameworks are implemented in \texttt{C}/\texttt{C++}, where gradient calculations are commonly derived and maintained by hand. This makes codebases monolithic and brittle: every change to the likelihood, prior, or forward model can trigger substantial and error-prone re-derivations, slowing the practitioners and impeding extensions that push beyond power spectra to cosmological parameters.

In this work, we address these challenges by introducing \flinch{}, a flexible and performant FLI framework that shares the hierarchical Bayesian structure of \almanac{} but with a crucial architectural difference: it is written entirely in the \julia{} programming language~\citep{bezanson2012julia} and leverages automatic differentiation (AD) for gradient calculations. This design substantially improves flexibility and development speed. The AD engine automatically applies the chain rule to compute exact posterior gradients, bypassing the laborious and error-prone manual implementations required by other frameworks. This not only streamlines development but also makes the model straightforward to extend: since \flinch{} is embedded in the native \texttt{Julia} AD ecosystem, it is straightforward to incorporate other differentiable cosmological tools.
By interfacing with differentiable codes for CMB~\citep{Bonici_2024}, 3D clustering~\citep{Bonici:2025ltp}, and 2D LSS observables~\citep{Bonici:2022xlo,Ruiz2024LimberJack, Chiarenza:2024rgk}, Bayesian inference can be pushed directly to the level of cosmological parameters, enabling seamless propagation of map-level uncertainties into final cosmological constraints. In addition to AD-enabled modeling, \flinch{} tackles the intrinsic sampling challenge of FLI: we compare three different samplers, HMC~\citep{DUANE1987216}, No-U-Turn Sampler (NUTS)~\citep{HoffmanGelman2011}, and {Microcanonical Langevin Monte Carlo (MCLMC)~\citep{Robnik2022, Robnik:2023pgt} on the problem studied here, highlighting their mixing behavior, tuning requirements, and computational cost in high-dimensional posteriors.

The paper is structured as follows. In Sec.~\ref{sec:problem} we describe the statistical framework we are considering and the forward model we implemented. In Sec.~\ref{sec:AD} we briefly describe AD and specify the main rules we developed for our analyses. In Sec.~\ref{sec:samplers} we describe the samplers employed in this analysis. In Sec.~\ref{sec:results} we describe the results we have obtained, both for the power spectrum reconstruction and cosmological parameter inference. We then summarise our findings in Sec.~\ref{sec:conclusions}.

\section{The problem}
\label{sec:problem}

Cosmological observables are intrinsically defined on the celestial sphere, and many leading probes can be represented as scalar (spin-0) or spin-2 fields. Describing data in harmonic space on the sphere provides a unified language to combine probes and to model survey effects such as masking, pixelization, and instrumental beams. In the hierarchical, field-level setting we adopt, the ultimate goal is inference on cosmological parameters\footnote{In this work we focus on cosmological parameters, but in a more realistic scenario we would include nuisance parameters that are related to foregrounds, systematics, and modelling.}, $\mathcal{P}(\boldsymbol{\theta}
\,|\,\mathbf{d})$, starting from pixelized maps $\mathbf{d}$. Directly targeting this posterior is typically intractable; for this reason, a growing body of literature has been applying Simulation Based Inference (SBI) methods~\citep{doi:10.1073/pnas.1912789117}. Usually SBI relies on some compression schemes, either driven by standard summary statistics or learned low-dimensional embeddings~\citep{Makinen:2021nly}, and on the ability of creating random realizations of the considered observables (hence the name Implicit Likelihood Inference)~\citep{2019MNRAS.488.4440A,  DES:2024xij, DES:2024jgw, vonWietersheim-Kramsta:2024cks}.

Here, following the \almanac{} paradigm~\citep{Loureiro_2023,Sellentin_2023}, we introduce latent spherical-harmonic coefficients $\mathbf{a}$ and (hyper-)parameters that control their covariance, and perform inference in a Bayesian graph that links maps, latent fields, and power spectra. Extending this hierarchy to include a prior on cosmological parameters, with power spectra $C_\ell(\boldsymbol{\theta})$ supplied by a differentiable theory module, allows us to exploit the constraining power of field-level inference while introducing controlled model dependence at the level of the two-point statistics rather than a full dynamical gravity model (as in forward-modeling approaches like \texttt{BORG}). In this section we formalize this setup on the sphere and detail the associated likelihood and priors; we then describe how we lift the hierarchy to cosmological parameters using a differentiable emulator. 

\subsection{The Hierarchical Inference Model}
\label{subsec:inference_problem}

\begin{figure}[!htp]
    \begin{center}
    \large{
    \begin{tikzpicture}[node distance = 1.4cm, auto]

        \pgfdeclarelayer{background}
        \pgfdeclarelayer{foreground}
        \pgfsetlayers{background,main,foreground}

        \tikzstyle{prob}=[draw, thick, text centered, rounded corners, minimum height=1em, minimum width=1em, fill=red!60]
        \tikzstyle{var}=[draw, thick, text centered, circle, minimum height=1em, minimum width=1em, fill=white]

        \begin{scope}[yshift = -2.8cm]
            \node[prob](prior){$\pi(\mathbf{C})$};
            \node[var](Cls)[below of= prior]{$\mathbf{C}$};
            \node[prob](Palms)[below of= Cls]{$\mathcal{G}(\mathbf{a} | \mathbf{C})$};
            \plate[inner sep=.2cm, yshift=.02cm, dashed] {plate1} {(Cls) (Palms)} {$\ell$};
            \node[var](alms)[below of= Palms]{$\mathbf{a}$};
            \node[prob](like)[below of= alms]{$\mathcal{L}(\mathbf{d} | \mathbf{a}, \mathbf{N})$};
            \node[var](noise)[left of= alms]{$\mathbf{N}$};
            \node[var](data)[below of= like]{$\mathbf{d}$};
            
            \path [draw, line width=0.7pt, arrows={-latex}] (prior) -- (Cls);
            \path [draw, line width=0.7pt, arrows={-latex}] (Cls) -- (Palms);
            \path [draw, line width=0.7pt, arrows={-latex}] (Palms) -- (alms);
            \path [draw, line width=0.7pt, arrows={-latex}] (alms) -- (like);
            \path [draw, line width=0.7pt, arrows={-latex}] (noise) -- (like);
            \path [draw, line width=0.7pt, arrows={-latex}] (like) -- (data);
        \end{scope}

        \begin{scope}[xshift=4cm]
            \node[prob](prior2){$\pi(\boldsymbol \theta )$};
            \node[var](thetas)[below of= prior2]{$\boldsymbol \theta$};
            \node[prob](deltaCls)[below of= thetas]{$\delta_{\rm D}\left(\mathbf{C} - \mathbf{C}\left(\boldsymbol \theta\right)\right)$};
            \node[var](Cls2)[below of= deltaCls]{$\mathbf{C}$};
            \node[prob](Palms2)[below of= Cls2]{$\mathcal{G}(\mathbf{a} | \mathbf{C})$};
            \plate[inner sep=.2cm, yshift=.02cm, dashed] {plate2} {(Cls2) (Palms2)} {$\ell$};
            \node[var](alms2)[below of= Palms2]{$\mathbf{a}$};
            \node[prob](like2)[below of= alms2]{$\mathcal{L}(\mathbf{d} | \mathbf{a}, \mathbf{N})$};
            \node[var](noise2)[left of= alms2]{$\mathbf{N}$};
            \node[var](data2)[below of= like2]{$\mathbf{d}$};
            
            \path [draw, line width=0.7pt, arrows={-latex}] (prior2) -- (thetas);
            \path [draw, line width=0.7pt, arrows={-latex}] (thetas) -- (deltaCls);
            \path [draw, line width=0.7pt, arrows={-latex}] (deltaCls) -- (Cls2);
            \path [draw, line width=0.7pt, arrows={-latex}] (Cls2) -- (Palms2);
            \path [draw, line width=0.7pt, arrows={-latex}] (Palms2) -- (alms2);
            \path [draw, line width=0.7pt, arrows={-latex}] (alms2) -- (like2);
            \path [draw, line width=0.7pt, arrows={-latex}] (noise2) -- (like2);
            \path [draw, line width=0.7pt, arrows={-latex}] (like2) -- (data2);
        \end{scope}

    \end{tikzpicture}
    }
    \end{center}
    \caption{Pictorial representation for our Bayesian hierarchical models. Left: baseline model following~\cite{Loureiro_2023}. A prior on the angular power spectra, $\pi(\mathbf{C})$, specifies the statistics of the latent field. Given these spectra, latent spherical-harmonic coefficients $\mathbf{a}$ are drawn from a zero-mean Gaussian with covariance set by $\mathbf{C}$. The observed data $\mathbf{d}$ are then generated from the likelihood $\mathcal{L}(\mathbf{d}|\mathbf{a},\mathbf{N})$, which combines the latent field with an additive noise component described by the covariance $\mathbf{N}$. Right: extended model introducing parameter vectors $\boldsymbol{\theta}$ with prior $\pi(\boldsymbol{\theta})$. The power spectra are no longer free variables but are tied to the parameters through the deterministic mapping $\mathbf{C}(\boldsymbol{\theta})$, implemented by a Dirac constraint. This reparameterization shifts the prior information from $\mathbf{C}$ to $\boldsymbol{\theta}$ while leaving the latent field layer $\mathbf{a}$ and the likelihood unchanged, enabling direct inference of the parameters from the data within the same generative structure.}
    \label{Fig:DAGs}
\end{figure}
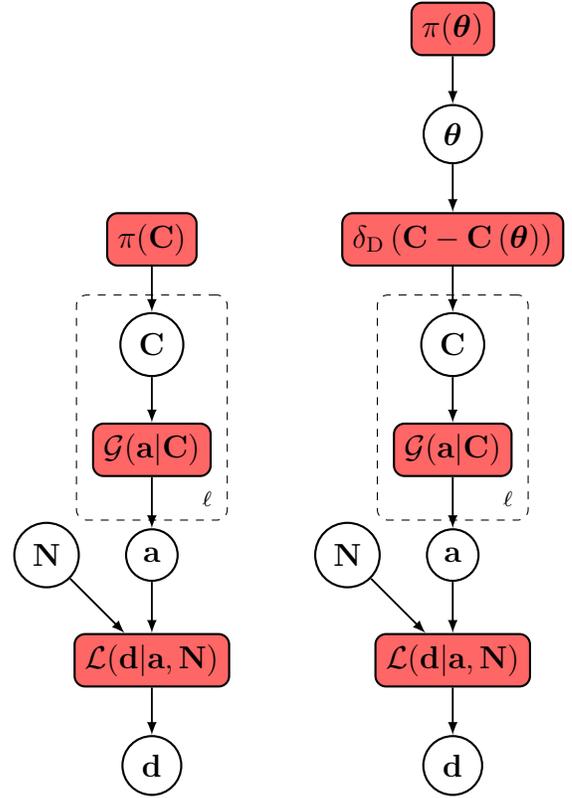

The optimal representation for fields on the full sky, where the flat-sky approximation fails, is a decomposition in spherical harmonics~\citep{Hu:1997hp}. For spin-0 fields we employ standard spherical harmonics; for higher-spin fields, spin-$s$ harmonics provide the appropriate generalization. The numerical spherical harmonic transform (SHT) for both scalar and spin-2 fields is implemented as a linear operator, enabling fast and parallel evaluation (see Sec.~\ref{sec:AD_SHT} for details on the differentiable SHT). Denoting by $\mathbf{a}$ the vector of harmonic coefficients of a spin-$s$ field and by $\mathbf{p}$ the corresponding pixel values, we write
\begin{equation}\label{eq:sht}
    \mathbf{p} = \mathbf{Y}\,\mathbf{a} \,,
\end{equation}
where $\mathbf{Y}$ contains the spin-$s$ spherical-harmonic basis functions ${}_{s}\mathcal{Y}_{\ell m}$. Two fields with harmonic coefficients ${}_{s}a^{(i)}_{\ell m}$ and ${}_{s}a^{(j)}_{\ell m}$ have self- and cross-spectra $C^{ij}_\ell$ defined by
\begin{equation}\label{eq:relaz_alm_Cl}
    \big\langle {}_{s}a^{(i)}_{\ell m}\; {}_{s}a^{(j)\,*}_{\ell' m'} \big\rangle
    = \delta_{\ell\ell'}\,\delta_{m m'}\,C^{ij}_{\ell} \,,
\end{equation}
with $i,j\in\{1,2\}$. This relation underpins the hierarchical link between fields and their power spectra in our inference model.

Let $\mathbf{d}$ denote the observed (masked) map. We model it as a noisy version of a latent realization $\mathbf{Y}\mathbf{a}$:
\begin{equation}
    \mathbf{d} = \mathbf{R}\,\mathbf{Y}\,\mathbf{a} + \mathbf{n} \,,
\end{equation}
where $\mathbf{R}$ is a linear response encoding pixelization and beam (defined below), and $\mathbf{n}\sim\mathcal{N}(\mathbf{0},\mathbf{N})$ captures uncorrelated Gaussian noise on unmasked pixels; masked pixels are simply removed and do not contribute to the likelihood. The posterior over latent coefficients $\mathbf{a}$ and power-spectrum coefficients $\mathbf{C}\equiv\{C_\ell\}$ then reads
\begin{equation}\label{eq:posterior}
    \mathcal{P}(\mathbf{a},\mathbf{C}\mid \mathbf{d},\mathbf{N}) \propto
    \underbrace{\mathcal{L}(\mathbf{d}\mid \mathbf{a},\mathbf{N})}_{\text{likelihood}}\,
    \overbrace{\mathcal{G}(\mathbf{a}\mid \mathbf{C})}^{\text{field prior}}\mkern-45mu
    \underbrace{\pi(\mathbf{C})}_{\text{power-spectrum prior}} \,,
\end{equation}
with Gaussian likelihood
\begin{equation}\label{eq:a_C_likelihood}
    \mathcal{L}(\mathbf{d}\mid \mathbf{a},\mathbf{N})
    \propto \exp\!\left[-\tfrac{1}{2}\big(\mathbf{d}-\mathbf{R}\mathbf{Y}\mathbf{a}\big)^\top
    \mathbf{N}^{-1}\big(\mathbf{d}-\mathbf{R}\mathbf{Y}\mathbf{a}\big)\right] \,,
\end{equation}
computed only on unmasked pixels \citep{2004PhRvD..70h3511W, 2009ApJ...697..258J}. Although masked regions are not directly constrained by the likelihood, large-scale modes couple information across the sky, enabling partial reconstruction beneath the mask in a field-level framework (as already demonstrated in \cite{Loureiro_2023, Sellentin_2023}). This hierarchical model is graphically represented in Fig.~\ref{Fig:DAGs}.

We include the dominant observational effects following \cite{sullivan2024}: the pixel-window function $P_\ell$ and the instrument beam $B_\ell$. Pixelization averages the sky within pixels, damping small-scale power; a common approximation is
\begin{equation}
    P_\ell = \mathrm{sinc}\!\left(\frac{\ell\,\theta_\mathrm{p}}{2\pi}\right) \,,\qquad
    \theta_\mathrm{p} = \sqrt{\frac{4\pi}{n_\mathrm{pix}}}\,,
\end{equation}
while a Gaussian\footnote{In this work we do not consider more complex beams that are usually employed in CMB analyses~\citep{Wandelt:2001gp, 2010ApJS..190..267P, Galloway:2022rhr}.} beam of width $\sigma$ implies
\begin{equation}
    B_\ell = \exp\!\bigg[-\frac{\ell(\ell+1)\sigma^2}{2}\bigg]\,.
\end{equation}
These effects rescale the true spectrum as $C^{\mathrm{obs}}_\ell = P_\ell^2 B_\ell^2 C_\ell$, and are applied at the map level via the response $\mathbf{R}$ so that the forward model is $\mathbf{R}\mathbf{Y}\mathbf{a}$.

For the field prior, we assume $\mathbf{a}$ is mean-zero with covariance set by $\mathbf{C}$. Reality of the map imposes $a_{\ell 0}\in\mathbb{R}$, while $a_{\ell m}$ for $m\neq 0$ are complex with independent real and imaginary parts. Preserving
$\langle a_{\ell m} a_{\ell' m'}^{*}\rangle = C_\ell\,\delta_{\ell\ell'}\delta_{m m'}$
requires sampling $\Re(a_{\ell m})$ and $\Im(a_{\ell m})$ with variance $C_\ell/2$ for $m\neq 0$, and variance $C_\ell$ for $m=0$:
\begin{equation}
\begin{split}
    \langle a_{\ell m}a_{\ell' m'}^{*}\rangle
    &=
    \langle \Re(a_{\ell m})\,\Re(a_{\ell' m'})\rangle
    + \langle \Im(a_{\ell m})\,\Im(a_{\ell' m'})\rangle \\
    &= \tfrac{C_\ell}{2}\delta_{\ell\ell'}\delta_{m m'}
     + \tfrac{C_\ell}{2}\delta_{\ell\ell'}\delta_{m m'}
     = C_\ell\,\delta_{\ell\ell'}\delta_{m m'} \,.
\end{split}
\end{equation}
Equivalently, writing a unified product over $\bar{m}$ that indexes both true $m$ values and the real/imaginary parts, the prior factorizes as
\begin{equation}
    \mathcal{G}(\mathbf{a}\mid \mathbf{C})
    \propto \prod_{\ell,\bar{m}} C_\ell^{-1/2}\,
    \exp\!\left[-\tfrac{1}{2}\,a_{\ell\bar{m}}\,
    \epsilon_{\bar{m}}^{-2}\,C_\ell^{-1}\,a_{\ell\bar{m}}\right] ,
\end{equation}
with $\epsilon_{\bar{m}}=1$ for $\bar{m}=0$ and $\epsilon_{\bar{m}}=1/\sqrt{2}$ otherwise. This is the maximum-entropy prior for a given covariance and does not, by itself, enforce Gaussianity of the posterior maps~\citep{Sellentin:2015waz}; non-Gaussian features in $\mathbf{d}$ can be reflected in the inferred $\mathbf{a}$ through the likelihood.

For the power-spectrum prior $\pi(\mathbf{C})$ we will either adopt weakly informative choices or, when lifting the hierarchy to cosmological parameters, replace it by a deterministic mapping $C_\ell = C_\ell(\boldsymbol{\theta})$ supplied by a differentiable theory module.

\subsection{Inference of cosmological parameters}
In the second, central part of this work we lift the hierarchy from power spectra to cosmology, replacing the hyperparameters $C_\ell$ with a differentiable map $C_\ell(\boldsymbol{\theta})$. We demonstrate this on simulated CMB temperature anisotropy maps, using the emulator \capse{}~\citep{Bonici_2024} to provide a fast, differentiable mapping
\begin{equation}
\boldsymbol{\theta} \longmapsto C_{\ell}\big(\boldsymbol{\theta}\big)\,.
\label{eq:capse_map}
\end{equation}
We consider the standard six-parameter $\Lambda$CDM vector
$\boldsymbol{\theta}=\{\ln(10^{10}A_\mathrm{s}),\, n_\mathrm{s},\, H_0,\, \omega_\mathrm{b},\, \omega_\mathrm{c},\, \tau\}$,
and fix the optical depth $\tau$ to its \Planck~2018 best-fit value, as temperature anisotropies alone weakly constrain it. To avoid extrapolating beyond the emulator’s training domain, we adopt independent uniform priors within the ranges covered by \capse{}:
\begin{equation}
\pi(\theta_i)=\mathcal{U}\!\bigl[\theta_i^{\min},\,\theta_i^{\max}\bigr] \,,
\end{equation}
with bounds listed in Tab.~\ref{tab:capse_range}. This choice is deliberately uninformative while ensuring well-behaved emulator derivatives throughout the prior volume. Within our hierarchy, the spectrum prior becomes a delta functional,
$\pi(\mathbf{C}\mid \boldsymbol{\theta})=\delta_\mathrm{D}\!\big(\mathbf{C} - \mathbf{C}(\boldsymbol{\theta})\big)$,
so that inference proceeds jointly over $\mathbf{a}$ and $\boldsymbol{\theta}$ under the likelihood in Eq.~\eqref{eq:a_C_likelihood} and the field prior in Sec.~\ref{subsec:inference_problem}.

Schematically, this posterior can be described as:
\begin{equation}\label{eq:posterior_cosmo}
    \mathcal{P}(\mathbf{a},\boldsymbol{\theta}\mid \mathbf{d},\mathbf{N}) \propto
    \underbrace{\mathcal{L}(\mathbf{d}\mid \mathbf{a},\mathbf{N})}_{\text{likelihood}}\,
    \overbrace{\mathcal{G}(\mathbf{a}\mid \mathbf{C}(\boldsymbol{\theta}))}^{\text{field prior}}\mkern-38mu
    \underbrace{\pi(\boldsymbol{\theta})}_{\text{cosmological prior}}\mkern-38mu.
\end{equation}
This hierarchical model is graphically represented in Fig.~\ref{Fig:DAGs}.

We highlight the similarity of this posterior and that of Eq.~\ref{eq:posterior}: the two share almost the same structure, with the same field-level prior and gaussian likelihood, the sole difference being represented by the priors on the highest level of the hierarchy, on the angular spectra and cosmological parameters, respectively, for the former and the latter approach. This showcases the flexibility of our framework, that can easily switch from a more agnostic approach to a more aggressive one that directly tackles cosmological parameters.

\begin{table}[t]
\centering
\begin{tabular}{lcc}
\toprule
Parameter & $\theta^{\min}$ & $\theta^{\max}$ \\
\midrule
$\ln(10^{10}A_{\mathrm s})$ & 2.5 & 3.5 \\
$n_{\mathrm s}$          & 0.88 & 1.06 \\
$H_0$[km/s/Mpc]              & 40. & 100. \\
$\omega_{\mathrm b}$ & 0.01933 & 0.02533 \\
$\omega_{\mathrm c}$ & 0.08 & 0.20 \\
\bottomrule
\end{tabular}
\caption{Prior ranges adopted for the cosmological parameters. All priors are uniform inside the interval and zero outside.}
\label{tab:capse_range}
\end{table}

We summarize here the practical prior choices and implementation details used in our cosmological runs. The uniform priors on $\boldsymbol{\theta}$ are bounded by the \capse{} training domain to maintain emulator fidelity and stable gradients. The adopted ranges are given in Tab.~\ref{tab:capse_range}. Within these bounds, we rely on the emulator’s differentiability to propagate parameter sensitivities through the hierarchy, enabling gradient-based samplers to explore the joint posterior efficiently. The latent-field prior and the observational response (pixel window and beam) are identical to those used in the power-spectrum hierarchy; noise and systematics are incorporated through $\mathbf{N}$ and the linear response $\mathbf{R}$. This setup ensures that field-level uncertainties and map-level systematics are coherently propagated into the final cosmological constraints.

\subsection{Reparameterization and posterior regularization}
\label{subsec:reparam}
In any sampling procedure, it is usually preferred to deal with the negative logarithm of the posterior itself:
\begin{equation}
    \psi(\mathbf{a},\mathbf{C}|\mathbf{d},\mathbf{N}) = -\log{\mathcal{P}(\mathbf{a},\mathbf{C}|\mathbf{d},\mathbf{N})} \ ,
\end{equation}
or
\begin{equation}
    \psi(\mathbf{a},\boldsymbol{\theta}|\mathbf{d},\mathbf{N}) = -\log{\mathcal{P}(\mathbf{a},\boldsymbol{\theta}|\mathbf{d},\mathbf{N})} \ .
\end{equation}
However, the sampling for this class of hierarchical Bayesian problems can be very challenging since the density distribution in the parameter space typically exhibits a funnel geometry~\citep{feba456f-8373-3f4e-b93a-903e0caf3a0f}. The difficulty arises because different parts of the funnel require dramatically different step sizes for adequate exploration, so a fixed-step sampler fails to identify a single typical scale and mixes poorly unless the step size is adapted dynamically or the parameterization is transformed to undo the funnel’s curvature~\citep{Papaspiliopoulos, Millea:2017fyd, Loureiro_2023}.
For single scalar fields, the prior distribution of the power spectrum coefficients is a simple uniform distribution over the positive real numbers. A better parametrization, which avoids this cut-off at zero, is to take $K_\ell = \log{C_\ell}$ (as done in \citealt{2008MNRAS.389.1284T}). In this way, the prior would be a uniform distribution over the entire real axis. However, we decided for an alternative parametrization which is able not only to avoid the cut-off, but it also limits the extension over the real axis, regularizing this parameter subspace. We transformed the $C_\ell$'s into a normally distributed variable, using the inverse of the cumulative distributions.
A fundamental property of any probability distribution is to be invariant under any change of coordinates. Hence, the final expression of the posterior has to be multiplied by the Jacobian of the reparameterization ($J_{C_\ell \rightarrow K_\ell}$) in order to guarantee this invariance:
\begin{equation}
    J_{C_\ell \rightarrow K_\ell}=\left|\frac{\mathrm{d}C_\ell}{\mathrm{d}K_\ell}\right| \propto \prod_\ell {\exp{\left[-\frac{K_\ell^{2}}{2}\right]}} \ .
\end{equation}

In the most general case of multiple fields, the \Cls{} become covariance matrices rather than scalars, which makes the parameter space geometry significantly more challenging. In such situations, a suitable reparameterization, such as working with the diagonal-log Cholesky factors of $C_\ell$, as shown in \cite{Loureiro_2023}, provides a more efficient and well-behaved sampling space. They also noted how the ratio $x_{\ell \bar{m}}=a_{\ell \bar{m}}/\sqrt{C_\ell}$ has always a unit variance; then, switching to the $x$ variables can make the geometry of the parameter space even more uniform. 
Finally, we also performed a regularization of the likelihood term: dividing the data map $\mathbf{d}$ and the sampled map $\mathbf{R}\mathbf{Y}\mathbf{a}$ by the square root of the noise matrix, the likelihood simplifies to a unit-variance gaussian distribution.

Directly sampling the cosmological parameters $\boldsymbol{\theta}$ for inference is inefficient, as their variances span several orders of magnitude and they show strong degeneracies. To mitigate this, we reparameterize the cosmological parameter block using an estimate of their covariance matrix, $\mathbf \Sigma_{\theta}$, inferred from the reconstructed $C_{\ell}$ chains in a procedure similar to post-processing. 

We estimate $\mathbf \Sigma_{\theta}$ as follows. First, we run our model-agnostic analysis to obtain the posterior distribution for the $C_{\ell}$ spectra. We then approximate this posterior as a multivariate normal distribution\footnote{Although this Gaussian approximation is not strictly exact, since the large-scale $C_{\ell}$ distributions deviate from normality~\citep{Hamimeche:2008ai, carrongaussian, Oehl:2024gbm}, this does not impact our goal. Since the covariance matrix is only used as a preconditioner, any deviation from the optimal preconditioner merely reduces sampling efficiency without introducing bias into the final parameter inference.}, estimating its mean and covariance directly from the chain samples. Using this approximate likelihood, we perform a standard summary-statistics analysis to infer the cosmological parameters. The resulting chains yield an estimate of $\mathbf \Sigma_{\theta}$.

Let now $\bm \Lambda$ be the matrix square root of $\mathbf \Sigma_{\theta}$ ($\mathbf \Sigma_{\theta}=\bm\Lambda\bm\Lambda^{\mathsf T}$) and define the transformations
\begin{equation}
\begin{split}
    \widetilde{\boldsymbol{\theta}} = \bm \Lambda^{-1}\,\boldsymbol \theta \\\boldsymbol \theta=\boldsymbol \Lambda\,\widetilde{\boldsymbol \theta} \, , 
\end{split} 
\label{eq:whiten_theta}
\end{equation}
then $\operatorname{Cov}(\widetilde{\theta}_i,\widetilde{\theta}_j)\simeq\delta_{ij}$ and so the transformed space is nearly isotropic, greatly improving the efficiency of the samplers.
Although this two–stage approach does not exploit the full joint distribution $\mathcal P(\mathbf a,\boldsymbol{\theta}|\mathbf{d})$, it provides a quick  estimate of the cosmological covariance matrix which is enough for our goal.

\subsection{Data simulation}
\label{subsec:data_sim}
Dealing with full-sky fields, both in pixel and harmonic coefficient spaces, is a non-trivial task. We used a \julia{} version of the well known algorithm \healpix{} by \cite{Gorski_2005}, which implements the original hierarchical pixelization of the sphere in equal-area pixels and includes the \texttt{ducc} algorithm to perform a fast spherical harmonic transform (an optimized version of \texttt{Libsharp2} algorithm by \cite{Reinecke_2013}). In particular, we opted for \healpixmpi{} by \cite{Bianchi2024}, which is an MPI-parallel implementation of the main functionalities of \healpix{}, allowing for high-performance spherical harmonic transform.
The pixel-space data are stored in a \texttt{map} object, which can handle both scalar and spin-2 fields. A map is characterized by a resolution parameter, $n_\mathrm{side}$, which is directly related to the total number of pixels used to describe the map itself, ($n_\mathrm{pix} = 12\,n_\mathrm{side}^2$).
In the same way, the complex spherical harmonic coefficients are stored in an \texttt{alm} object, a vector ordered by $m$. However, since the inference is performed separately on the real and imaginary part, firstly we store the harmonic coefficient components in a distinct \texttt{alm} object and then transform it back to the \healpixmpi{} interface.
We simulated the CMB measured maps at a given resolution $n_\mathrm{side}$, based on a fiducial set of power spectrum coefficients, obtained with \capse{}. Then, using a built-in \healpix{} function, we generated a random realization map from these fiducial coefficients, added a Gaussian noise with a noise level of $\sim12 \, \mu \mathrm{K}$ for $n_\mathrm{side}=512$ (properly rescaled for different resolutions), and finally applied a WMAP survey mask~\citep{Bennett_2013, Hinshaw_2013}).

\section{Automatic Differentiation}
\label{sec:AD}
High dimensionality is the primary source of complexity in field-level inference: even modest resolutions already imply tens of thousands of latent degrees of freedom, and realistic surveys produce maps at $n_\mathrm{side}=1024$ or $2048$ (or higher), pushing the number of parameters into the millions to tens of millions once latent fields, hyperparameters, and nuisance terms are included. In this regime, effective sampling typically requires gradient-based methods, most notably HMC and its modern variants, which simulate Hamiltonian dynamics to use posterior gradients for long-distance, geometry-aware proposals, yielding far better scaling than random-walk schemes in high dimensions~ \citep{Betancourt2017}. The catch is that this shifts the computational burden from proposing states to evaluating exact and efficient gradients with respect to millions of variables, making gradient computation a central algorithmic bottleneck for both flexibility and performance.

Classical numerical differentiation via finite differences (FD) scales linearly with the number of parameters and accumulates truncation and round-off errors, rendering it impractical at field-level dimensionalities. Fully analytical gradients, as in \almanac{}, avoid FD pathologies but sacrifice flexibility: any change to parameterizations, priors, or components of the pipeline requires re-deriving and re-implementing gradient expressions by hand. This rigidity is further exacerbated when pushing inference beyond power spectra to cosmological parameters via a hierarchical model. In standard pipelines, key map-level quantities are produced by numerical Boltzmann solvers as functions of cosmological parameters; propagating hand-written derivatives through these black-box numerical stages is cumbersome and, in practice, has not been achieved at scale. Consequently, maintaining closed-form gradients across all layers of the hierarchy quickly becomes untenable, especially when extending the model or changing parameterizations.

These considerations motivate an AD approach. AD computes exact derivatives by decomposing the program into elementary differentiable operations and applying the chain rule compositionally\footnote{For this reason it is also dubbed algorithmic differentiation.}. Two computational modes exist. In forward mode, directional derivatives are propagated from inputs to outputs; its cost scales with the number of inputs, which is prohibitive for our high-dimensional latent spaces. In backward (reverse) mode, sensitivities are pulled back from outputs to inputs; its cost scales with the number of outputs (often small for scalar objectives like log posteriors), making it the appropriate regime for field-level inference. Formally, if $x$ maps to $y$, the adjoint variable is defined as
\begin{equation}
\bar{x} \equiv \frac{\mathrm{d}y}{\mathrm{d}x},
\end{equation}
and adjoints together with the chain rule drive reverse-mode accumulation, yielding all input sensitivities from a single reverse pass. The geometric underpinnings of these push-forward (forward mode) and pull-back (reverse mode) operators are discussed in \cite{betancourt2018}, which provides a differential-geometric view of AD and its relation to statistical computations.

In practice, our gradients are computed with \zygote{}~\citep{innes2019}, a reverse-mode AD tool optimized for \julia{} and well-suited to large-scale, differentiable scientific programs. AD’s key advantage is flexibility: when the model, priors, or parameterization change, only the corresponding adjoint rules (if any custom rules are needed) or code paths must be updated, rather than re-deriving entire gradient expressions by hand. The trade-off is a modest overhead: reverse mode first evaluates the function and then executes a reverse pass to accumulate gradients. With careful implementation, wall-clock time can approach the ideal regime where computing the function and its gradient is roughly twice the cost of the function alone~\citep{griewank2008evaluating}. Finally, because the pipeline is fully differentiable, coupling to differentiable cosmology components (e.g., emulators or differentiable Boltzmann solvers~\citep{Hahn:2023nvb, Sletmoen:2025fro}) enables direct inference in cosmological parameter space with gradients flowing from maps to parameters, making end-to-end, field-to-parameter inference operational within the same HMC-based framework.

\subsection{AD Spherical Harmonic Transform}
\label{sec:AD_SHT}
To demonstrate how AD works in practice for a relevant example in our analysis, we consider the spherical harmonic transform. Following the notation of \healpix, we refer to the equation for transforming spherical harmonic coefficients to a map as the \almtomap{} function, and use \adjalmtomap{} to denote the adjoint operation:
\begin{equation}
    \mathbf{a} = \mathbf{Y}^{\mathrm{T}}\mathbf{p} \ .
\end{equation}
(Attention: here \textit{adjoint} in \adjalmtomap{} simply refers to the fact that the operation is performed with the adjoint of the matrix $\mathbf{Y}$.)
The pseudo-code snippet in Fig.~\ref{fig:ad_sht_snippet} implements the adjoint rule for \almtomap\footnote{See \cite{2024JCoPh.51013109P} for a differentiable SHT implemented in \texttt{jax}.}.

\begin{figure}[h!]
    \centering
    \includegraphics[width=\columnwidth]{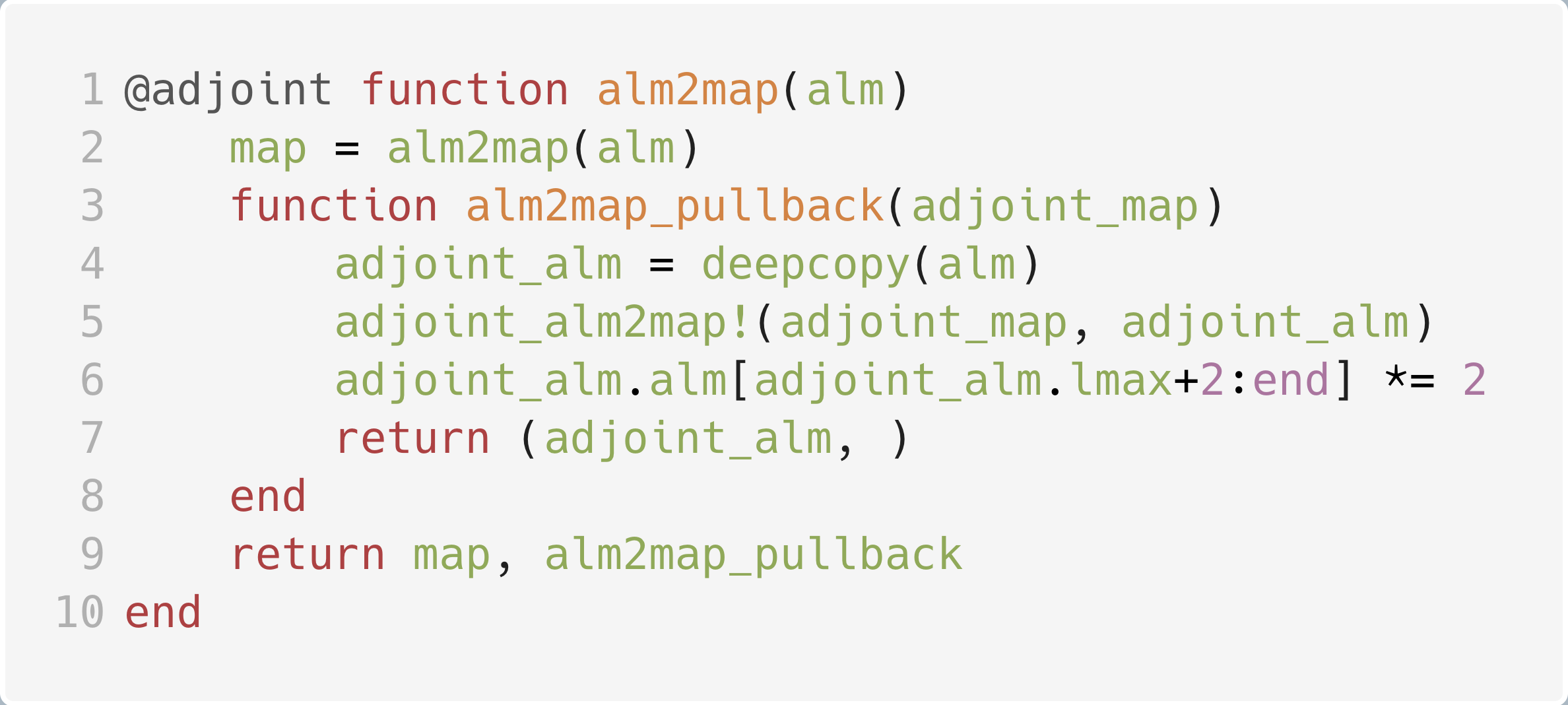}
    \caption{Example of an adjoint-aware wrapper for a spherical-harmonic transform. The forward pass converts a set of harmonic coefficients \texttt{alm} into a \texttt{map}, while the pullback reconstructs the gradient in harmonic space by reusing the same routine and applying the parity fix for the highest modes. This illustrates how AD support can be added with only a few lines of code, once the mathematical rules are derived.}
    \label{fig:ad_sht_snippet}
\end{figure}

The main component of this adjoint implementation is the \almtomapPB{} function. This function takes an adjoint from the output space (represented by \texttt{adjoint\_map}, which is a map-like object) and pulls it back to the input space (producing \texttt{adjoint\_alm}, an alm-like object). The \texttt{adjoint\_alm} is closely related to the actual gradient of the function with respect to the spherical harmonic coefficients and can be passed as input to the next step (or, more precisely, the previous step in the reverse mode) of the AD computation.
At line 6 in the snippet, the \texttt{adjoint\_alm} is modified by scaling the coefficients beyond the maximum multipole index, $\ell_\mathrm{max}$, by a factor of 2. This scaling is necessary to account for specific symmetries in the SHT and to ensure that the gradient is computed correctly. Specifically, \almtomap{} is a special case where the map represents a real-valued field, which imposes reality conditions on the spherical harmonic coefficients: $a_{\ell -m}=a^{*}_{\ell m}$ and $\Im(a_{\ell 0})=0$. Due to this symmetry, \healpix{} only needs to store the coefficients with $m\geq0$. However, both positive and negative $m$-modes are involved in the computation of the function and its gradient. In the gradient calculation, each entry of \texttt{adjoint\_alm} at fixed $\ell$ and $m$ contains information about the derivative with respect to both $a_{\ell m}$ and $a_{\ell -m}$, which are equal due to symmetry.

When benchmarking our forward model and likelihood evaluation, we find that the SHTs constitute the primary computational bottleneck, with wall-times consistent with those reported in~\cite{Bianchi2024}. With our hand-crafted differentiation rules, a single gradient evaluation, which includes both the forward and backward passes, takes about 2.1 times as long as the corresponding forward pass.

\section{Samplers}
\label{sec:samplers}
In this work, we use a range of gradient-based samplers to efficiently explore the high-dimensional posterior distribution that we introduced in Sec.~\ref{sec:problem}. The employed algorithms are a standard implementation of the HMC sampler, NUTS, MCLMC, and Pathfinder. In the reminder of this section, we briefly describe these algorithms.

\subsection{Hamiltonian Monte Carlo}  
\label{sec:HMC}
Hamiltonian Monte Carlo is a Markov Chain Monte Carlo (MCMC) method that uses the Hamiltonian dynamics of a system to efficiently sample from complex, high-dimensional posterior distributions. Originally developed to address the inefficiencies of random-walk-based MCMC algorithms, HMC uses gradients of the posterior distribution to propose new points that maintain high acceptance rates and explore the parameter space efficiently.
The method introduces auxiliary momentum variables $\bm{p}$, associated with each parameter variable $\bm{x}$, to form a combined state $(\bm{x},\bm{p})$ that evolves according to Hamiltonian dynamics. The total Hamiltonian function is defined as the sum of a potential energy and kinetic energy:
\begin{equation}
    H(\bm{x}, \bm{p}) = U(\bm{x}) + K(\bm{p}) \ .
\end{equation} 
$U(\bm{x})$ is chosen as the negative logarithm of the target distribution, such that $U(\bm{x})= - \log{\mathcal{P}(\bm{x})}$, and $K(\bm{p})$ is typically a quadratic function of $\bm{p}$, often defined as $K(\bm{p})=\frac{1}{2}\bm{p}^\mathrm{T}\mathbf{M}^{-1}\bm{p}$, where $\mathbf{M}$ is a mass matrix that can be adjusted to improve sampling efficiency.

The probability density for a given state $(\bm{x},\bm{p})$ is then given by: 
\begin{equation} 
    \pi(\bm{x},\bm{p}) \propto \exp[-H(\bm{x}, \bm{p})] = \exp[-U(\bm{x}) - K(\bm{p})] \, , 
\end{equation} 
which allows us to sample both position and momentum simultaneously, evolving the system along trajectories defined by Hamilton's equations.
In the standard HMC algorithm, a new proposal state is generated by numerically integrating Hamiltonian dynamics for a fixed number of steps $n_\mathrm{steps}$ with a step size $\epsilon$. Both these parameters are crucial to the performance of the algorithm, as too large a step size or too few steps can result in low acceptance rates, while too small a step size or too many steps can slow down exploration and increase computational cost. For this reason, although it is highly effective for many applications, it can be challenging to tune.
The No-U-Turn Sampler is an extension of HMC that aims at automatically adjusting the number of integration steps to improve efficiency, removing the need to manually tune this parameter. Proposed by \cite{HoffmanGelman2011}, NUTS addresses one of the primary limitations of standard HMC by dynamically determining when to stop a trajectory based on the geometry of the posterior distribution. The algorithm builds a tree of candidate states by simulating the Hamiltonian dynamics both forward and backward in time. The tree structure allows the sampler to double the number of integration steps at each iteration, expanding the trajectory until a \textit{U-turn} is detected. When the sampler starts to turn back on its path, further steps would lead to redundant exploration. In this way, NUTS avoids unnecessary computation and ensures efficient exploration of the parameter space. The NUTS algorithm also includes an adaptive mechanism for tuning the step size during an initial warm-up phase, further enhancing the efficiency and robustness of the sampling process. This adaptation is performed using dual averaging, which adjusts $\epsilon$ to achieve a target acceptance rate~\citep{Nesterov2009}.
Finally, both standard HMC and NUTS perform a Metropolis-Hastings acceptance step to decide whether accepting the proposed step or not.
In our implementation, we make use of the \julia{} package \texttt{AdvancedHMC.jl} from the \texttt{Turing.jl} ecosystem~\citep{10.1145/3711897}, which allows us to have a single framework for different choices of HMC samplers. 

\subsection{MicroCanonical Langevin Monte Carlo}
\label{sec:MCLMC}
MicroCanonical Langevin Monte Carlo offers an alternative sampling paradigm to canonical methods like HMC by operating within the microcanonical ensemble, where the total energy of the system is conserved~\citep{Robnik2022, Robnik:2023pgt, 2025arXiv250301707R}. This approach is designed to overcome some of the limitations of traditional samplers, particularly in high-dimensional spaces. Instead of sampling from a canonical distribution where states have varying energies, MCLMC confines the system's evolution to a constant-energy surface in the phase space.

The core idea of MCLMC is to define a modified Hamiltonian such that the marginal distribution of the positions on this constant-energy surface corresponds to the desired target distribution. This is achieved by employing isokinetic Langevin dynamics, which keeps the kinetic energy of the system constant. The dynamics are described by a system of stochastic differential equations (SDEs) that are dissipation-free, meaning they do not require the balance between fluctuation and dissipation that characterizes canonical samplers~\citep{Kubo:1966fyg}.

The continuous-time evolution of a particle's position $\bm{x}$ and velocity $\bm{u}$ in MCLMC is given by the following SDEs:
\begin{equation}
    \begin{cases}
        \mathrm{d}\bm{x} = \bm{u} \mathrm{d}t \\
        \mathrm{d}\bm{u} = \mathbf{P}(\bm{u}) \left( -\frac{\nabla U(\bm{x})}{d-1} \mathrm{d}t + \eta \mathrm{d}\bm{w} \right)
    \end{cases}
\end{equation}
where $U(\bm{x}) = -\log p(\bm{x})$ is the potential energy, $d$ is the dimension of the parameter space, $\bm{w}$ is a Wiener process representing Gaussian white noise, and $\eta$ is a parameter controlling the strength of the stochastic perturbation. The operator $\mathbf{P}(\bm{u}) = \mathbf{I} - \bm{u}\bm{u}^\mathrm{T}$ is a projection matrix that ensures the velocity vector $\bm{u}$ remains on the surface of a unit sphere, thus keeping the kinetic energy constant. The stochastic term $\eta \mathbf{P}(\bm{u})\mathrm{d}\bm{w}$ introduces noise perpendicular to the velocity, which ensures some degree of ergodicity by allowing the sampler to explore the entire constant-energy surface, a crucial property for correct sampling.

A key feature of MCLMC is that it is an unadjusted sampler, meaning it does not employ a Metropolis-Hastings acceptance step. Every state generated by the integration of the dynamics is accepted as a new sample. This let the sampler takes larger steps in parameter space and can lead to significant computational savings. However, this unadjusted nature introduces a numerical bias that is dependent on the integration step size $\epsilon$. The bias can be controlled by monitoring the Energy Error Variance Per Dimension (EEVPD) and keeping it below a certain threshold. This provides a practical way to manage the trade-off between sampling efficiency and bias~\citep{2025arXiv250301707R, Simon-Onfroy:2025ziw}.

MCLMC has shown promising results, particularly for high-dimensional problems where it can significantly outperform HMC and its variants like NUTS~\citep{bayer2023, Simon-Onfroy:2025ziw, Chen:2025wdy}. For instance, in applications like lattice field theory, MCLMC has demonstrated speedups of over an order of magnitude compared to HMC~\citep{Robnik:2023pgt}. While MCLMC introduces a bias, its favorable scaling and efficiency make it a powerful tool for tackling challenging sampling problems in modern scientific computing. A recent development, the Metropolis-Adjusted Microcanonical Sampler (MAMS), introduces a Metropolis-Hastings step to MCLMC to produce asymptotically unbiased samples while retaining much of its efficiency~\citep{2025arXiv250301707R}.

\subsection{Pathfinder} 
\label{sec:pathfinder}
Pathfinder~\citep{zhang_2022} is a quasi-Newton variational inference (VI) method that builds a sequence of multivariate normal approximations along an L-BFGS optimization trajectory targeting the log posterior, using the optimizer’s inverse-Hessian estimates to capture local curvature efficiently~\citep{LiuNocedal1989}. At each iteration, Pathfinder defines a Gaussian approximation whose covariance is given by the (negative) inverse-Hessian estimate from L-BFGS and scores these candidates via a Monte Carlo estimate of the evidence lower bound (ELBO), ultimately selecting the approximation with the lowest estimated Kullback–Leibler (KL) divergence to the true posterior. In its multi-path variant, multiple independent runs are combined by importance resampling to form a mixture-of-Gaussians approximation, improving robustness to non-normality, minor modes, plateaus, and saddle points.

Empirically, Pathfinder produces approximate posterior draws that often match or exceed the quality of ADVI and approach those from short dynamic HMC (measured by 1-Wasserstein distance), while requiring one to two orders of magnitude fewer log-density and gradient evaluations, with larger gains on challenging posteriors~\citep{zhang_2022}. Because Pathfinder targets the posterior directly (rather than optimizing a stochastic ELBO objective) and evaluates ELBOs “embarrassingly” in parallel along the path, it can deliver rapid approximate samples suitable for initialization of MCMC samplers. In particular, Pathfinder is effective as a warm-start for MCMC methods, reducing burn-in by starting near typical, high-probability regions; Pareto-smoothed importance diagnostics can help assess reliability of resampled draws for initialization versus direct inference~\citep{JMLR:v25:19-556}.

In our analyses, we use Pathfinder\footnote{Specifically, we have used \texttt{Pathfinder.jl}~\citep{seth_axen_2025_16818742}.} solely as an initializer for HMC, NUTS, and MCLMC, rather than as a stand-alone estimator, to avoid non-ergodicity risks and potential failures to fully explore complex, multimodal posteriors. We observe the largest practical benefit for MCLMC, approximately a five-fold reduction in tuning iterations to reach comparable convergence, whereas HMC and NUTS exhibit more modest improvements.

\section{Results}
\label{sec:results}
We organize the results into two parts. First, we investigate the reconstruction problem, recovering the map and its angular power spectrum, and assess convergence using a suite of diagnostics. Convergence statistics are computed intra-chain, i.e., across multiple independent chains initialized differently for each sampler. Concretely, for HMC we ran \(n_{\mathrm{hmc}}\) chains with 10{,}000 tuning steps followed by 10{,}000 samples; for NUTS, \(n_{\mathrm{nuts}}\) chains with 1{,}500 tuning steps and 1{,}000 samples; and for MCLMC, \(n_{\mathrm{MCLMC}}\) chains with 2{,}000 tuning steps and 10{,}000 samples, thinned by a factor of 2. We also compare sampler efficiency to address the practical viability of these computationally intensive analyses, and a detailed discussion of these aspects, together with all the relevant figures, can be found in Appendix~\ref{app:convergence_perf}. Second, we move to cosmological parameter inference, where the field-level likelihood is used to derive posteriors on a set of cosmological parameters. In both parts, we perform a detailed comparison against a standard pseudo-\Cls{} pipeline based on \namaster{}, thereby isolating the methodological impact on reconstructed spectra and on cosmological constraints. 

\subsection{Map and angular spectra reconstruction}
\label{sec:map_cls_reconstruction}
\begin{figure*}[t]
    \centering
    \begin{subfigure}{0.48\textwidth}
        \centering
        \includegraphics[width=\linewidth]{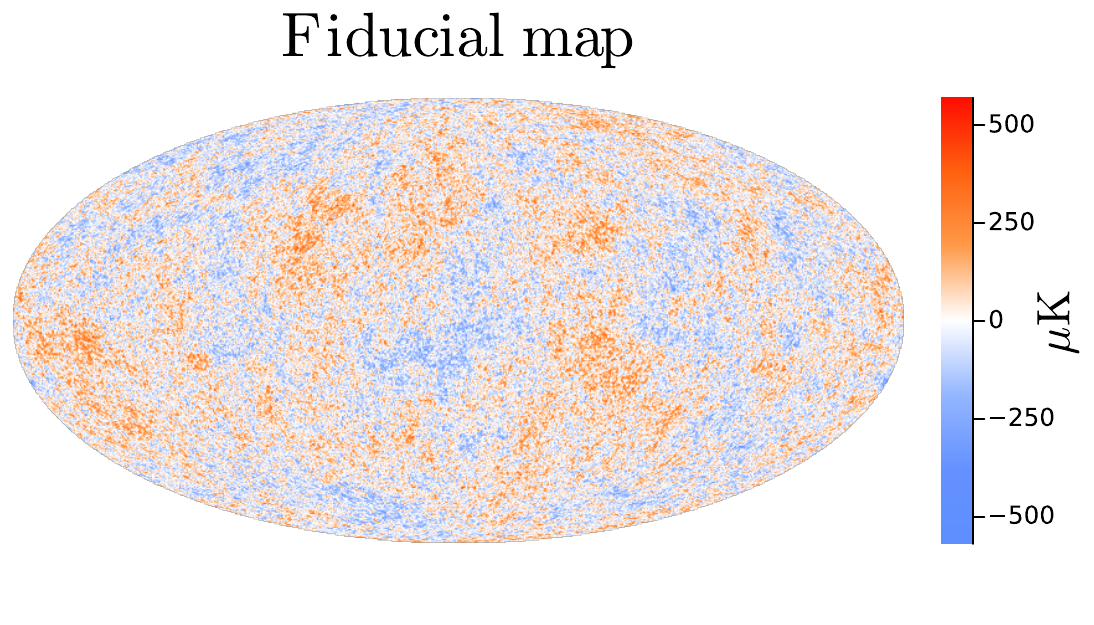}
        \caption{}
    \end{subfigure}
    \hfill
    \begin{subfigure}{0.48\textwidth}
        \centering
        \includegraphics[width=\linewidth]{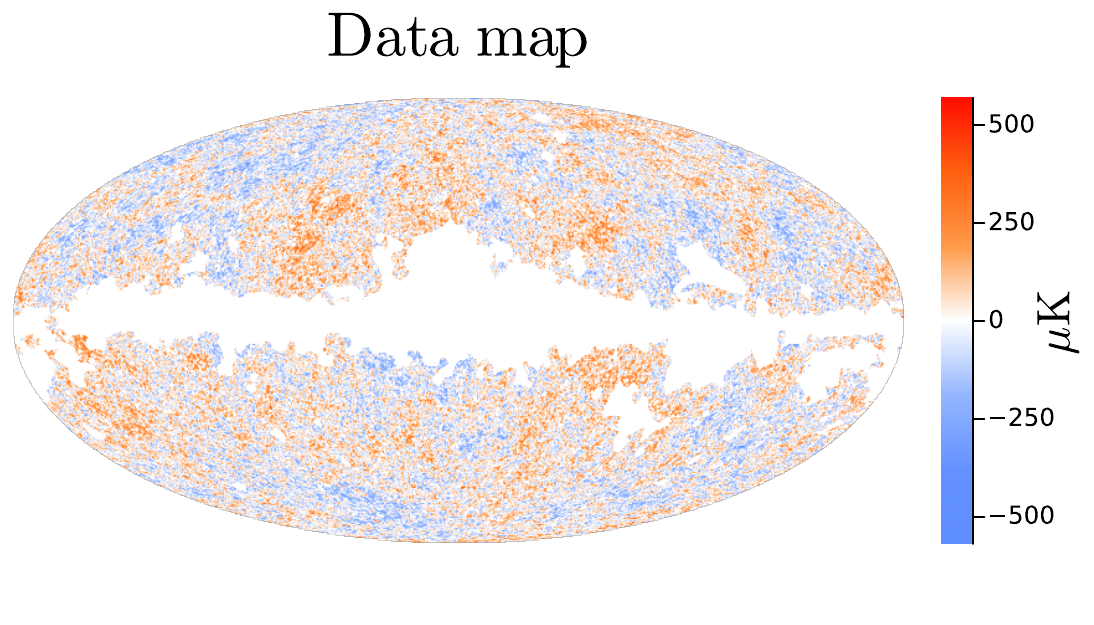}
        \caption{}
    \end{subfigure}

    \begin{subfigure}{0.48\textwidth}
        \centering
        \includegraphics[width=\linewidth]{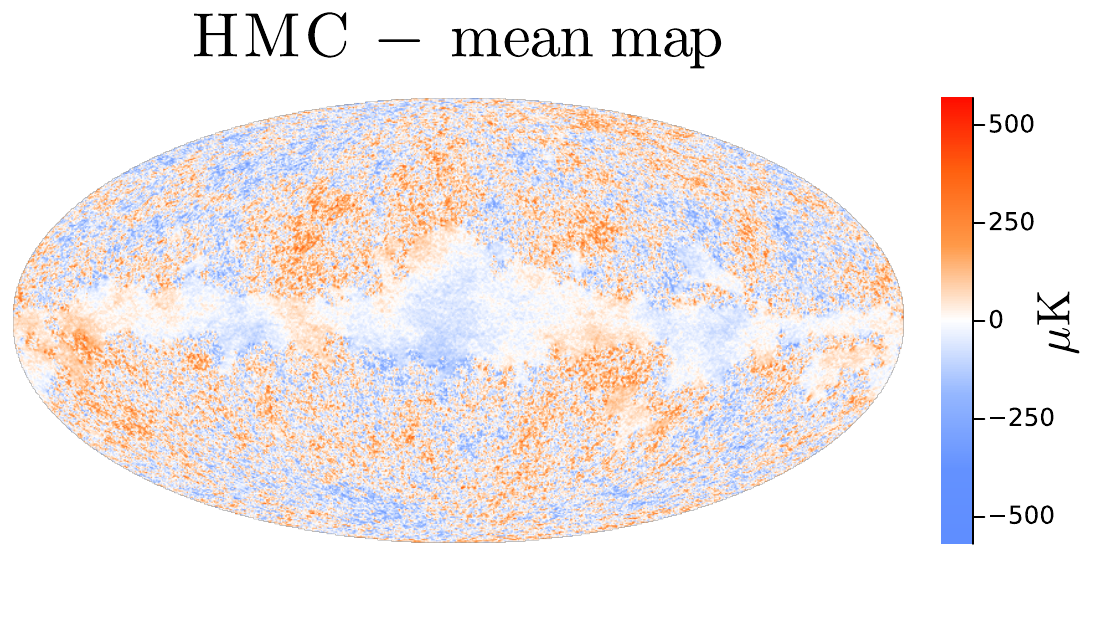}
        \caption{}
    \end{subfigure}
    \hfill
    \begin{subfigure}{0.48\textwidth}
        \centering
        \includegraphics[width=\linewidth]{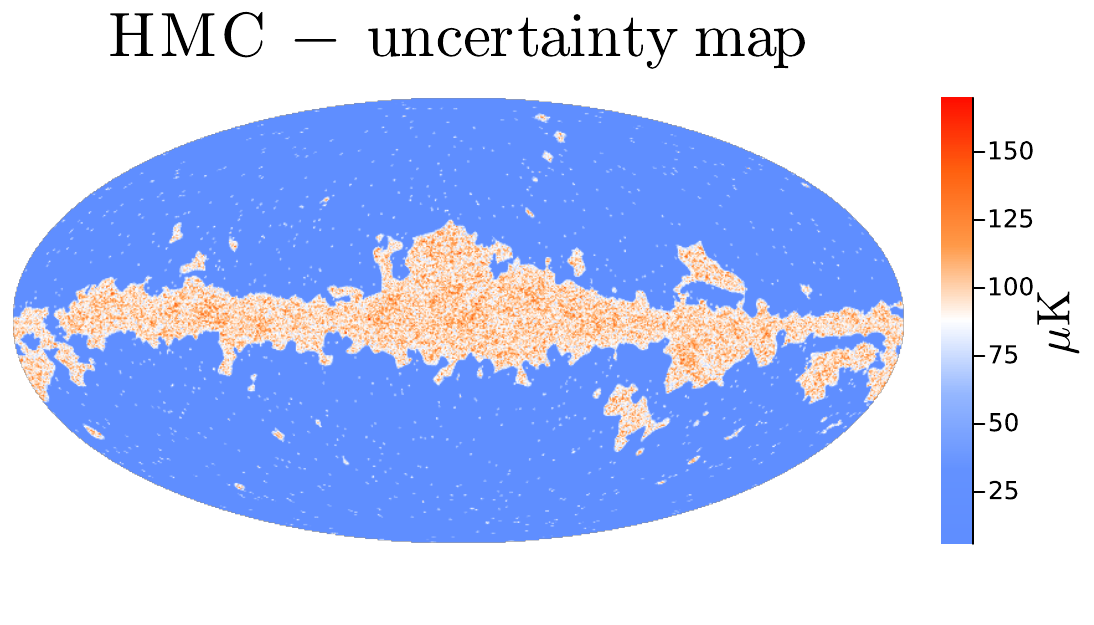}
        \caption{}
    \end{subfigure}

    \begin{subfigure}{0.48\textwidth}
        \centering
        \includegraphics[width=\linewidth]{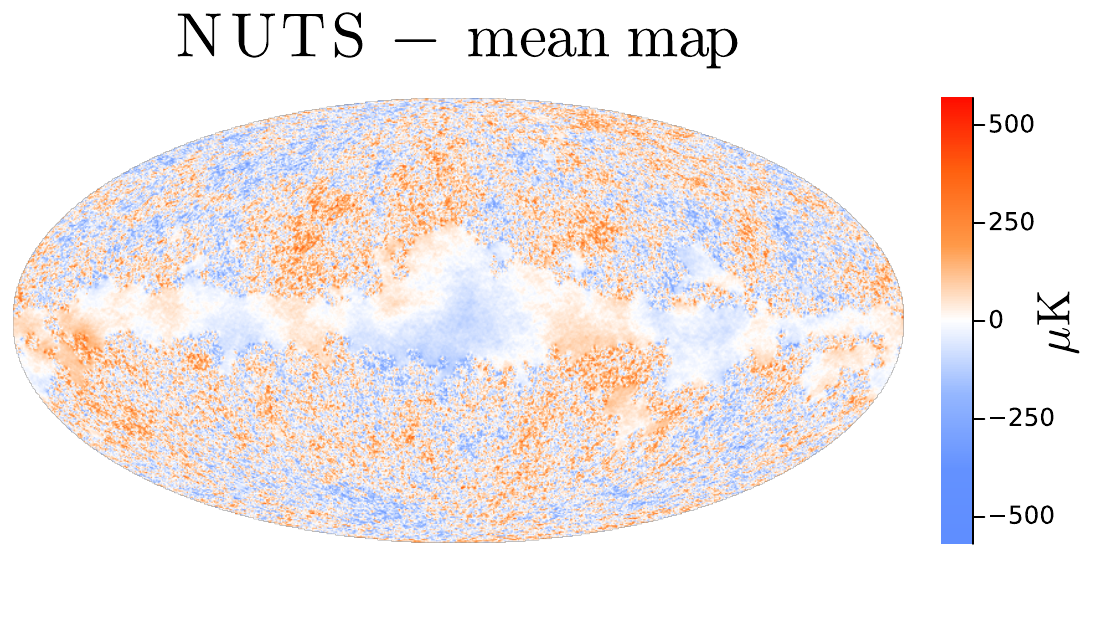}
        \caption{}
    \end{subfigure}
    \hfill
    \begin{subfigure}{0.48\textwidth}
        \centering
        \includegraphics[width=\linewidth]{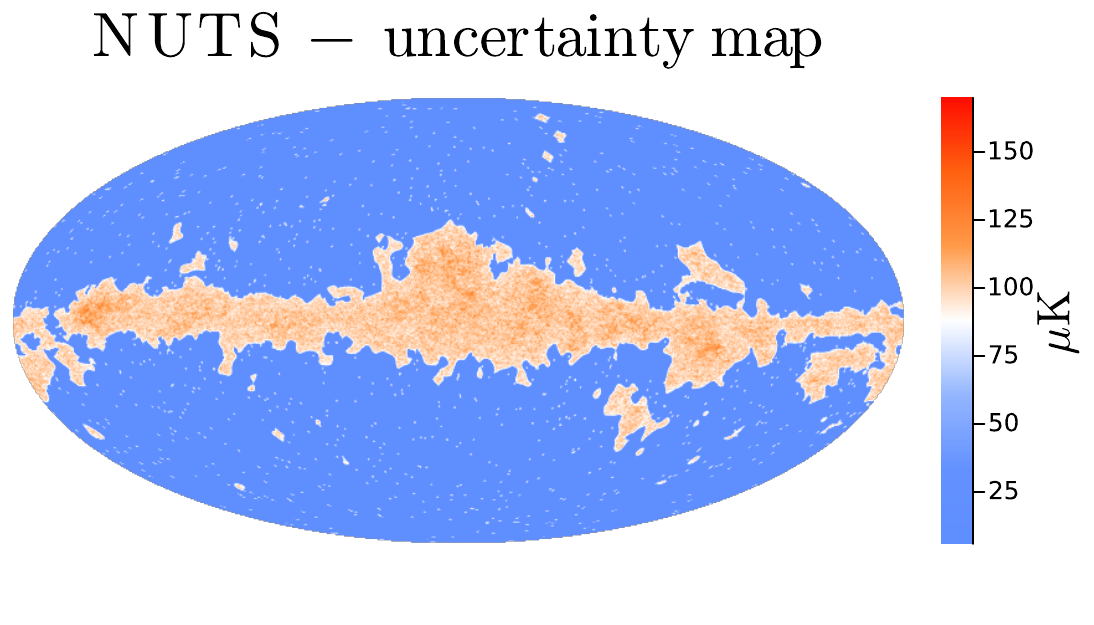}
        \caption{}
    \end{subfigure}

    \begin{subfigure}{0.48\textwidth}
        \centering
        \includegraphics[width=\linewidth]{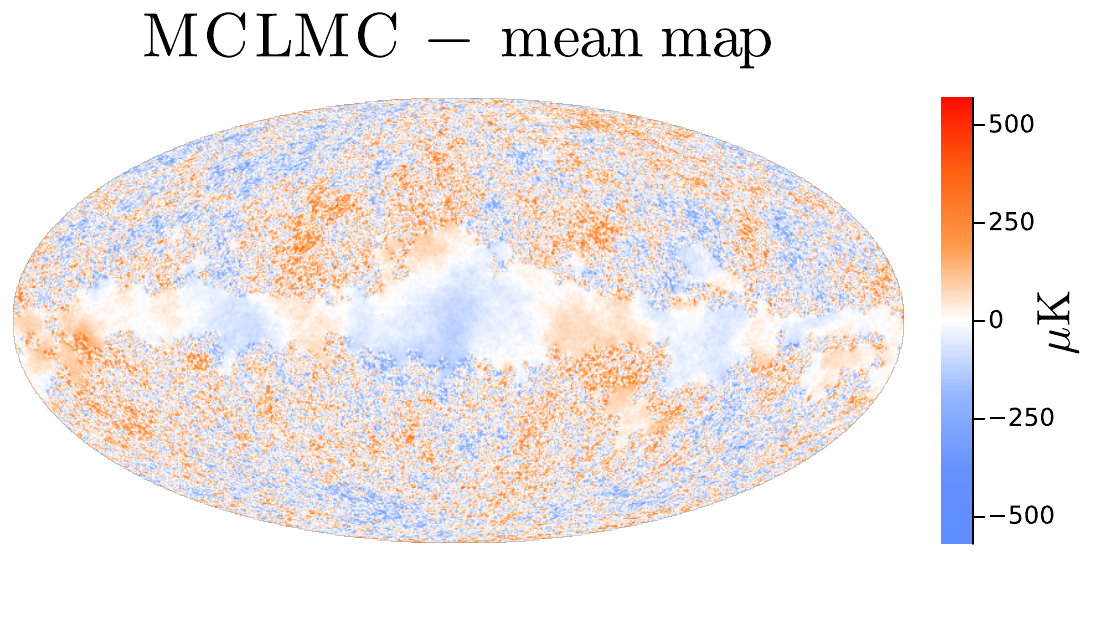}
        \caption{}
    \end{subfigure}
    \hfill
    \begin{subfigure}{0.48\textwidth}
        \centering
        \includegraphics[width=\linewidth]{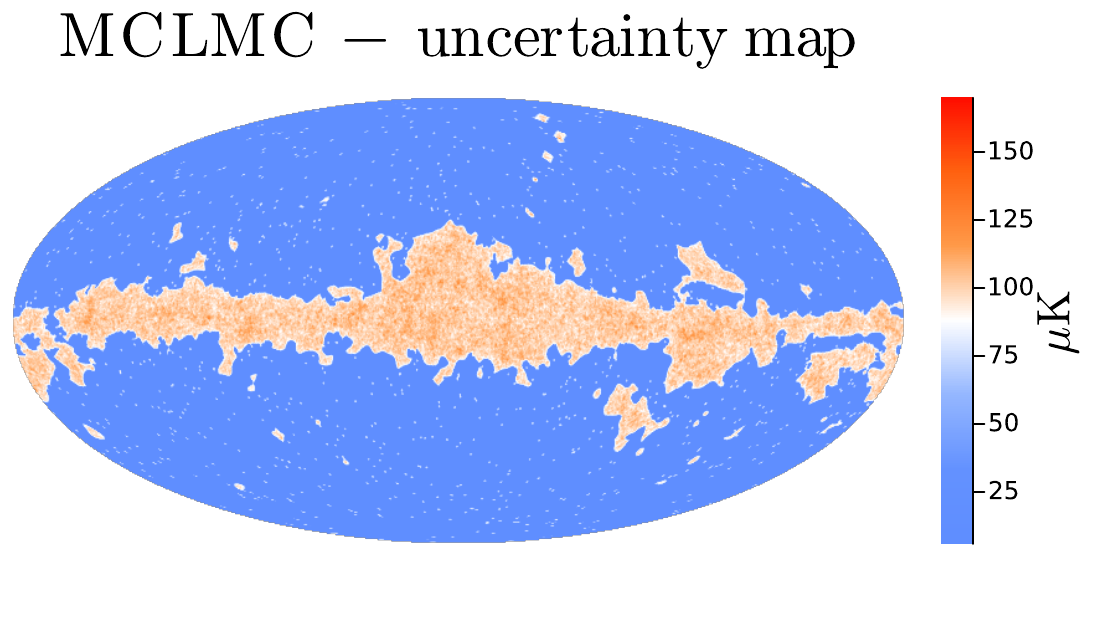}
        \caption{}
    \end{subfigure}

    \caption{Reconstruction of a masked, noisy field. Panels: (a) fiducial noiseless map; for each reconstruction method, we show the mean map and the per‑pixel standard deviation: HMC (c–d), NUTS (e–f), and MCLMC (g–h). All maps share the same color scale. The uncertainty panels highlight higher variance within the masked region, while the mean reconstructions recover the large-scale structure beneath the mask.}
    \label{fig:recon_std_maps}
\end{figure*}
In Fig.~\ref{fig:recon_std_maps}, we show the comparison between the noiseless realization of the map and its noisy, masked version, together with the mean reconstruction obtained from the chains for each sampler. Alongside, we also report, for each case, the corresponding standard deviation map derived from the chains.
The results are as expected: outside the masked region, the map reconstruction is highly accurate at all scales and also very precise, as indicated by the very low standard deviation. Inside the masked region, however, the field-level pipeline can only recover information on large scales, and the reconstruction is less precise, resulting in a significantly higher standard deviation. We recall that this is because large-scale modes, with wavelengths much larger than the masked region, can be partially recovered through correlations with unmasked areas, as their structure is constrained by the global mode coupling in the data. In contrast, small-scale modes under the mask are almost entirely unconstrained, as their fluctuations are dominated by local information that is absent in the observed data map. As a result, the reconstruction in these regions naturally recovers only the broad, large-scale features, while fine-scale structure is effectively lost. The reconstruction accuracy was assessed, as performed by the \almanac{} team, through the distribution of the pixel residuals, i.e., for each pixel \(p_i\) of the map and for each sampler \(s\) we evaluated the quantity 
\begin{equation}
\mathcal{R}_{s,i}=\frac{p_{i,s}^\mathrm{recon} - p_{i,s}^\mathrm{fid}}{\sigma_{i,s}^\mathrm{recon}} \,
\end{equation}
namely the residual between the reconstructed value of the map and the fiducial realization, normalized by the corresponding standard deviation. For a good reconstruction, we expect the residuals to follow a standard Gaussian distribution. Indeed, as shown in Fig.~\ref{fig:map_res}, the histograms of the three samplers almost perfectly overlap with the standard Gaussian curve (plotted in black as a reference).
In practice, however, HMC shows a slight deviation from the standard Gaussian, indicating a less optimal reconstruction. This is also apparent from the HMC reconstructed map, which, under the mask, appear noisier than those from the other two samplers, with correspondingly higher peaks in the standard deviation.

\begin{figure}
\centering
\includegraphics[width=\columnwidth]{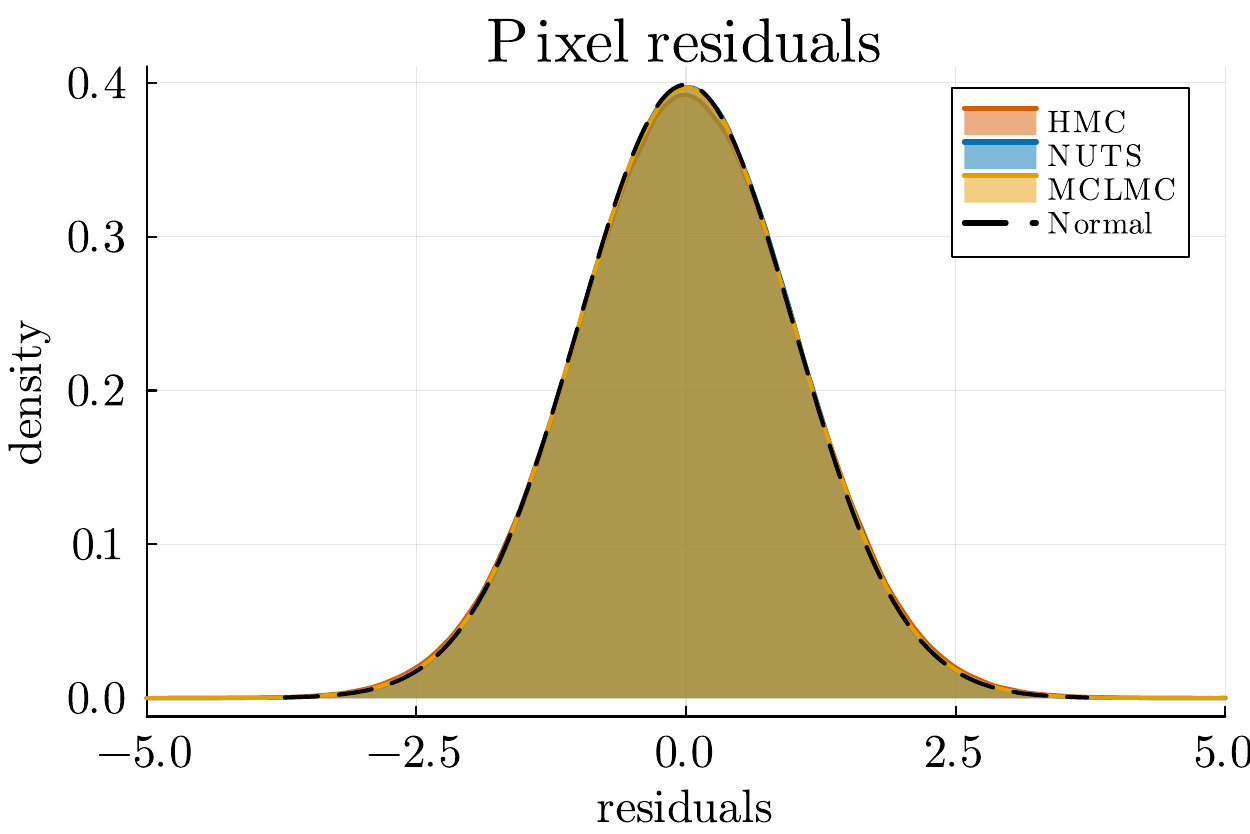}
\caption{Pixel residuals normalized by the recovered per-pixel uncertainty. For NUTS and MCLMC, the residual distributions nearly perfectly overlap a standard normal ($\mu=0$, $\sigma^2=1$), consistent with their high degree of convergence. In contrast, HMC shows a slight deviation from this behavior, reflecting its comparatively weaker convergence.}
\label{fig:map_res}
\end{figure}

\begin{figure}
\centering
\includegraphics[width=\columnwidth]{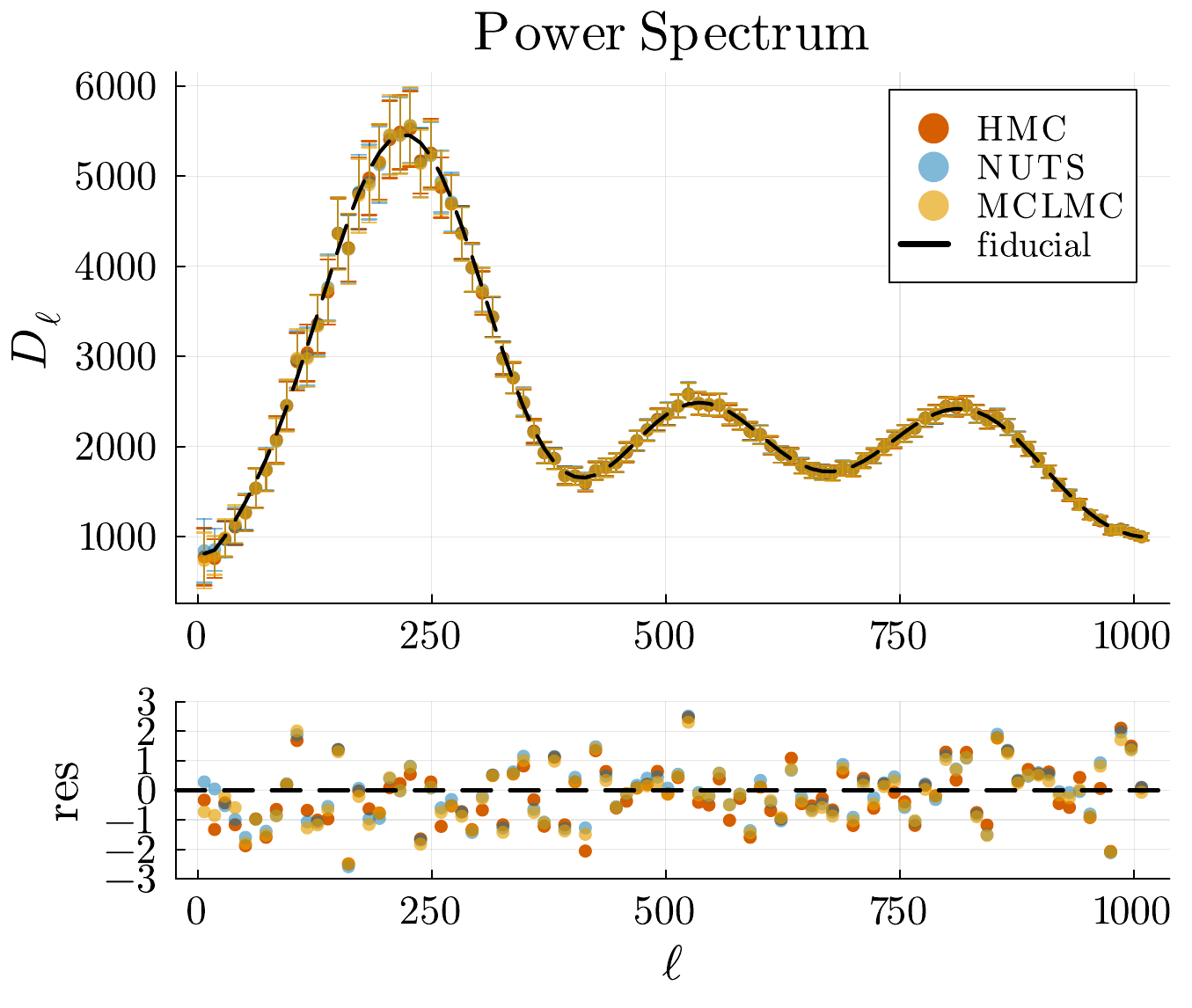}
\caption{Power spectrum comparison. The dashed curve shows the underlying input (fiducial) theory, while the markers display the reconstructed spectra with their $3\sigma$ standard deviations. The lower panel reports residuals normalized by their corresponding standard deviations, highlighting consistency with the fiducial model across scales. For clarity, we rebinned the entire \Cls{} range so that each bin contains 11 multipoles.}
\label{fig:power_spec}
\end{figure}

\begin{figure}[!htp]
\centering
\includegraphics[width=\columnwidth]{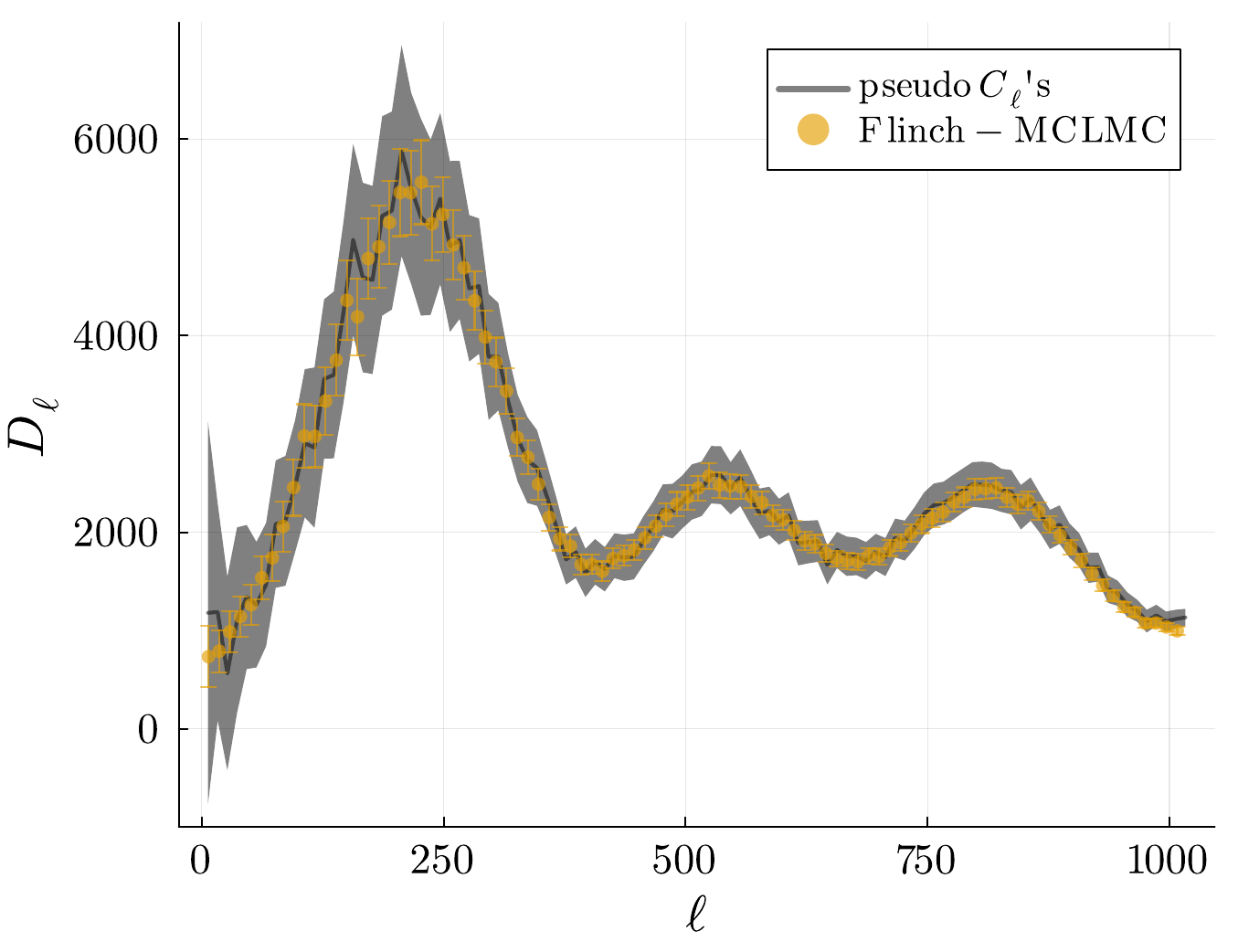}
\caption{Binned power spectra recovered by the three samplers (HMC, NUTS, MCLMC) and by the standard pseudo-$C_{\ell}$ pipeline implemented in \texttt{NaMaster}. The comparison highlights the sub-optimality of the pseudo-$C_{\ell}$ framework: its estimator exhibits substantially larger variance than the \flinch{} reconstructions across scales, demonstrating the efficiency gains from the field-level approach. For visual comparison, the error bars and the shaded area are shown at three times the standard deviation.}
\label{fig:flinch_vs_namaster}
\end{figure}

\begin{figure*}
\centering
\includegraphics[width=\textwidth]{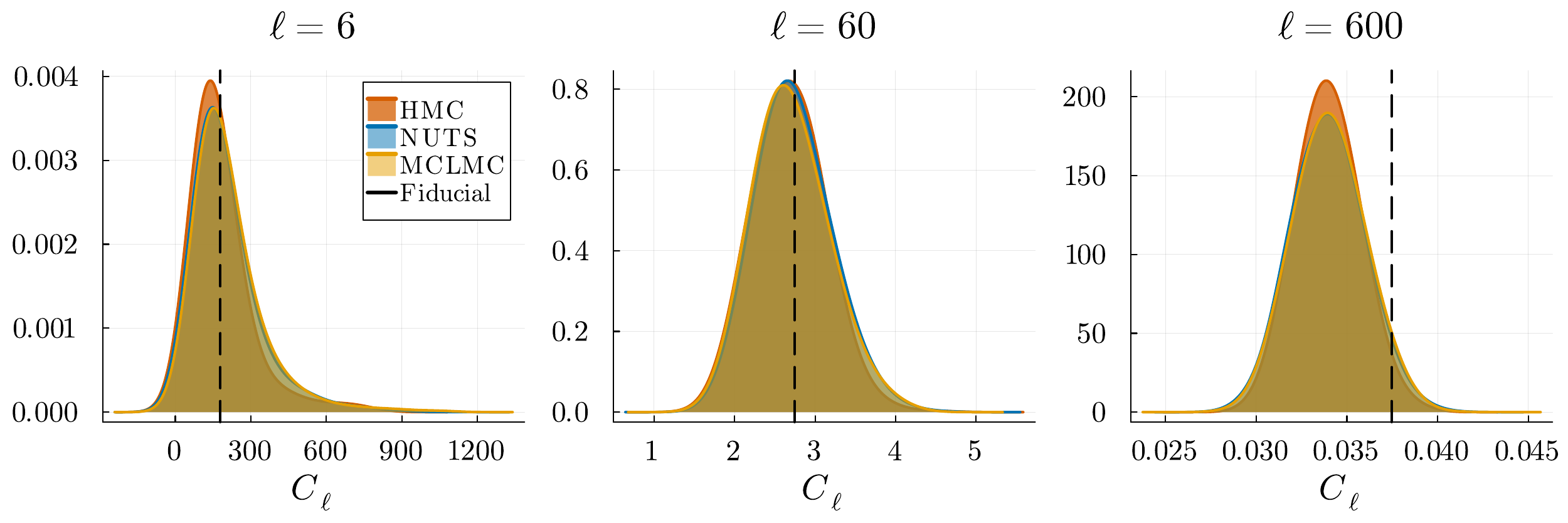}
\caption{Posterior distributions for three representative power-spectrum multipoles, $C_{\ell}$, selected to probe large ($\ell=6$), intermediate ($\ell=60$), and small ($\ell=600$) scales. Across all three cases, HMC, NUTS, and MCLMC produce mutually consistent posteriors. At the largest scales ($\ell=6$), the distribution departs noticeably from a Gaussian shape, whereas at intermediate and small scales the posteriors are well-approximated by Gaussians.}
\label{fig:Cl_posterior}
\end{figure*}

We now turn to the reconstruction of the multipoles of the angular power spectrum. 
Overall, Fig.~\ref{fig:power_spec} shows that the three samplers are capable of reconstructing the full power spectrum across the entire multipole range considered. The plotted values correspond to the medians of the chains, since at low \(\ell\) the distributions are not symmetric; the error bars correspond to 99.7\% confidence intervals (\(3\sigma\) interval). To obtain a clean and clear plot, we rebinned the entire \Cls{} range so that each bin contains 11 multipoles. In the bottom panel, we also show the residuals, computed as the difference between the reconstructed and fiducial values (black dashed curve), normalized by the standard deviation of the chain. Visual inspection shows an almost perfect consistency with a standard normal distribution. In Fig.~\ref{fig:Cl_posterior}, we also compare, for three example values of \(\ell\), the posterior distributions reconstructed by the three samplers. Once again, we find that the three samplers yield equivalent distributions, with the exception of HMC, which in some cases deviates from the mean reconstructed by NUTS and MCLMC due to suboptimal convergence. 
It is also clearly visible that for low multipoles, \(\ell=6\) in this case, all three samplers are able to recover the expected non-Gaussianity of the probability distribution. Indeed, since the \Cls{} are obtained from the square of the Gaussian-distributed \alm's, they follow a \(\chi^2\) distribution. However, as the multipole increases, more modes are combined, and by the central limit theorem the probability distribution tends to become Gaussian, as can already be seen at \(\ell=60\) and even more so at \(\ell=600\).
After establishing the samplers’ accuracy, we next assess their computational performance, as scalability to high-dimensional parameter spaces hinges on sampler efficiency. In our highest-dimensional configuration, MCLMC achieves a markedly superior convergence rate, being about 80 times more efficient than NUTS and roughly 800 times more efficient than standard HMC, underscoring the practical advantages of the microcanonical family for field-level inference at scale. More details on this comparison are given in Appendix~\ref{app:convergence_perf}.

To enable a direct, like-for-like assessment of performance and to compare the statistical FLI approach with the methodology routinely employed in cosmological analyses, we compared the spectra obtained with \flinch{} against those derived from the standard pseudo-\Cls{} pipeline, where both the pseudo-\Cls{} and their covariance are estimated with \namaster{}~(\citealt{Alonso:2018jzx, Garc_a_Garc_a_2019}). In the conventional workflow for masked-sky analyses, \namaster{} provides bias-corrected pseudo-\Cls{} as well as an estimate of their covariance matrix. However, this approach is intrinsically suboptimal, as extensively discussed in the literature, particularly when benchmarked against estimators that more fully exploit the data, such as Quadratic Maximum Likelihood (QML) methods~\citep{Efstathiou:2003dj, Seljak:2017rmr, Alonso:2018jzx, Garc_a_Garc_a_2019, Maraio:2022ywi}. This suboptimality manifests as a larger uncertainty on reconstructed spectra, as clearly illustrated in Fig.~\ref{fig:flinch_vs_namaster}, where, for the reconstruction case and using identical multipole binning, we display the power spectra inferred by \flinch{} with all three samplers alongside the \namaster{} pseudo-\Cls{}. To aid visual comparison, the error bars are shown at three times the standard deviation; the tighter constraints achieved by the field-level approach remain evident across the multipole range. 

With these results, which confirmed that our pipeline is capable of reconstructing the information contained in the map in order to accurately estimate our chosen summary statistic, we can now push the inference directly to the cosmological parameters level.\\

\subsection{Cosmological inference}
\label{sec:cosmo_inference}
\renewcommand{\arraystretch}{1.5} 
\begin{table*}
    \centering
    \begin{tabular}{lccccc}
        \toprule
         & \textbf{$\log{10^{10}A_\mathrm s}$} & \textbf{$n_\mathrm s$} & \textbf{$H_0$}[km/s/Mpc] & \textbf{$\omega_\mathrm b$} & \textbf{$\omega_\mathrm c$} \\
        \midrule
        \textbf{Fiducial} &  $3.0$  & $0.96$ & $67.0$ & $0.0225$ & $0.12$ \\
        \midrule
        \textbf{Flinch -- HMC} &  $2.97^{+0.14}_{-0.13}$ & $0.965^{+0.034}_{-0.033}   $ & $69^{+8}_{-8}              $  & $0.0217^{+0.0016}_{-0.0014}$  & $0.113^{+0.031}_{-0.028}   $ \\
        \textbf{Flinch -- NUTS} &  $2.96^{+0.14}_{-0.13}$ & $0.963^{+0.032}_{-0.034}$ & $69^{+8}_{-8}$  & $0.0216^{+0.0016}_{-0.0018}$ & $0.112^{+0.032}_{-0.027}$ \\
        \textbf{Flinch -- MCLMC} &  $2.98^{+0.13}_{-0.12}$ & $0.970^{+0.035}_{-0.033}$& $68^{+8}_{-8}              $  & $0.0219^{+0.0015}_{-0.0015}$ & $0.116^{+0.031}_{-0.027}$ \\
        \textbf{Pseudo-\(C_\ell\) -- NUTS} &  $3.03^{+0.20}_{-0.18}      $  & $0.968^{+0.051}_{-0.047}   $  & $66^{+10}_{-10}            $ & $0.0215^{+0.0022}_{-0.0020}$ & $0.125^{+0.050}_{-0.042}   $ \\
        \midrule
        \textbf{$\%$ improvement} &  $32\%$  & $33\%$ & $20\%$ & $19\%$ & $37\%$ \\
        \bottomrule
    \end{tabular}
    \caption{Comparison of cosmological parameter estimates obtained with different inference methods. The first row contains the fiducial values of cosmological parameters used to generate the stochastic realization. The second, third, and fourth rows show the obtained constraints on cosmological parameters. The final row compares the standard deviation obtained using \flinch{}-NUTS and \namaster{}. The results confirm that \texttt{Flinch.jl} yields systematically tighter constraints than the pseudo-$C_\ell$ analysis, with error bars reduced by about 20–40\%, depending on the parameter. This demonstrates that the field-level likelihood retains additional statistical information compared to the pseudo-$C_\ell$ approach, leading to sharper and more precise cosmological constraints.}
    \label{tab:cosmo_params}
\end{table*}
\renewcommand{\arraystretch}{1}
In this section, we present cosmological parameter inference with the field-level approach, which provides the most compelling test of the method by enabling the use of all information in the map without assuming a likelihood for intermediate summary statistics.
While this approach is not supported by the current \almanac{} implementation, it becomes a straightforward extension in our framework thanks to AD, which enables propagation of derivatives from the maps to cosmological parameters when coupled to a differentiable theory code~\citep{Campagne:2023ter, Piras2023CosmoPower, Bonici_2024, Hahn:2023nvb, Ruiz2024LimberJack, Balkenhol:2024sbv, Sletmoen:2025fro}.
Besides our approach, there are at least a couple of alternatives which do not rely on a fully differentiable model. First, one might train a normalizing flow to model the posterior density of the reconstructed \Cls{}, obtained from either \almanac{} or \flinch{}, and then use this learned density as an effective likelihood in combination with a standard Boltzmann solver or an emulator (\citeauthor{Nazli_inprep}, \textit{in prep}.). Second, it is possible to sample from such a class of models using a Gibbs sampler~\citep{Alsing:2016hkh, Paradiso:2022fky}.

We perform this analysis on a map generated as in the reconstruction-only setup but at a lower resolution, \(n_{\mathrm{side}}=256\), to better control computational cost and to allow multiple runs with different configurations. We perform a five-parameter inference for $\log{10^{10}A_{\mathrm{s}}}$, $n_{\mathrm{s}}$, $H_0$, $\omega_{\mathrm{b}}$, and $\omega_{\mathrm{c}}$, using all three samplers, which obtain remarkable agreement.

As shown previously in our power-spectrum reconstruction tests, the pseudo-\Cls{} approach implemented via \namaster{} is suboptimal for non-flat spectra, while \flinch{} achieves tighter, more faithful reconstructions when applied on the same data. We now assess the impact of this difference at the level of cosmology by directly comparing posterior constraints on cosmological parameters inferred with \flinch{} against those obtained from the standard pseudo-\Cls{} pipeline (pseudo-\Cls{} and covariances estimated with \namaster{}, combined with \capse{}'s theory prediction). As you can also see in Tab.~\ref{tab:cosmo_params}, \flinch{} still achieves error bars that are between 20\% and 40\% narrower, depending on the cosmological parameter.

Consistent with expectation, the parameter posteriors derived with \flinch{} are systematically tighter than those from the pseudo-\Cls{} analysis, confirming that the information retained by the field-level likelihood meaningfully sharpens cosmological constraints.

It is important to stress that the improvements we observe already emerge in the deliberately simple setting of a Gaussian random field, where only two-point statistics are informative; if we were to incorporate higher-order contributions both when generating and analyzing the data with an Edgeworth expansion~\citep{Sellentin:2017aii, Philcox:2021ukg}, we could thus expect more pronounced gains. Equally crucial is that these gains stem purely from methodological advances: by improving the statistical framework, we achieve tighter constraints under identical data, masks, and resolution. Looking ahead to the transition from Stage-III to Stage-IV surveys, this highlights that methodological progress should advance alongside hardware upgrades, so that enhanced statistical tools can maximize the scientific return of next-generation surveys.

\begin{figure*}[t]
\centering
\includegraphics[width=0.95\textwidth]{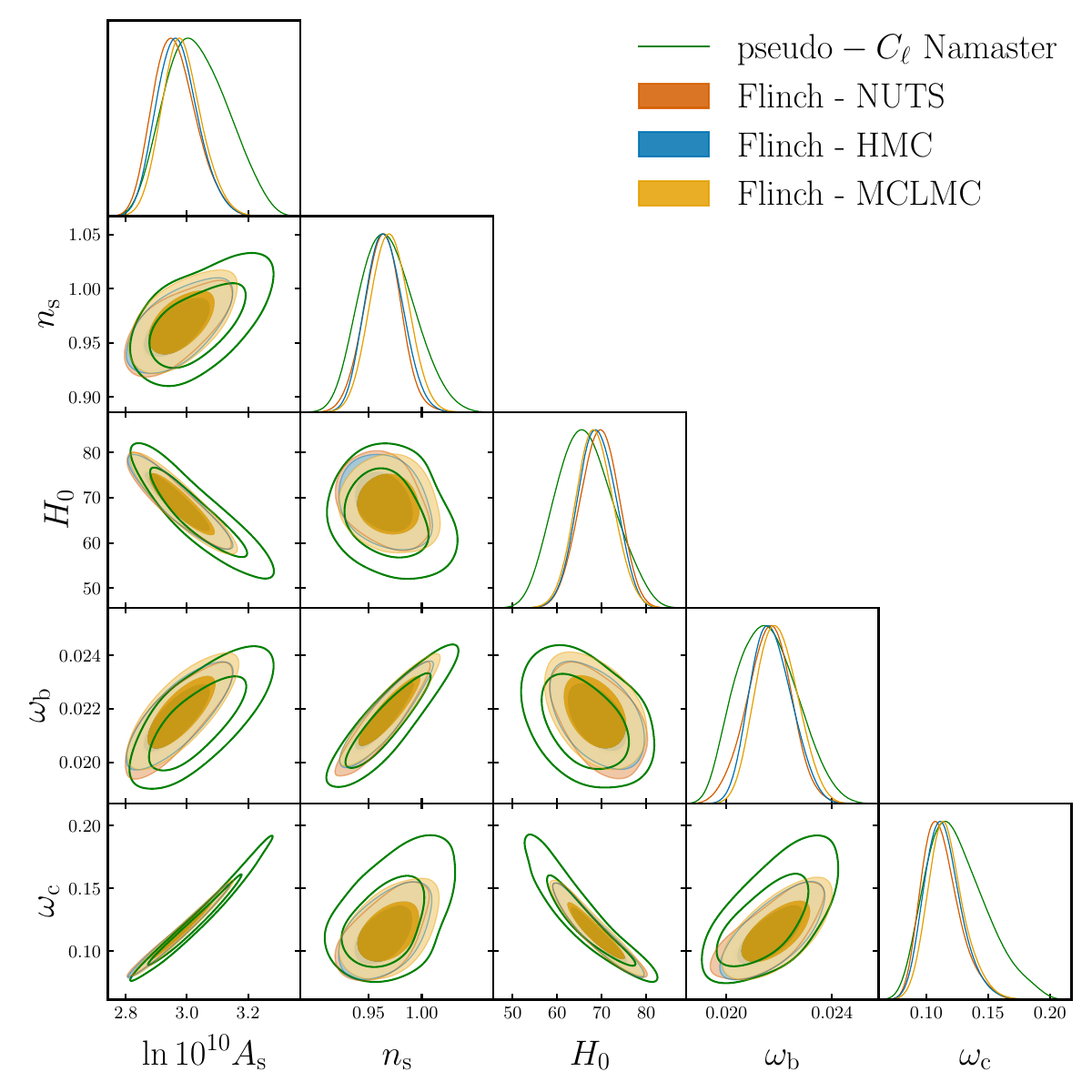}
\caption{Marginalized posterior contours for key cosmological parameters. For \flinch{} (HMC, NUTS), the inference hierarchy is pushed directly to the cosmological parameters, while the \texttt{NaMaster} pipeline employs a Gaussian likelihood with the \texttt{NaMaster} covariance. As in previous figures, the \texttt{NaMaster} approach is suboptimal: \flinch{} yields noticeably tighter constraints across parameters.}
\label{fig:allparams_corner}
\end{figure*}

\section{Conclusion}
\label{sec:conclusions}
In this work, we have presented \flinch, a new \julia{}-based, fully differentiable framework for field-level inference for data defined on the sphere. Building on the hierarchical Bayesian structure of previous approaches, but written entirely in \julia{} and leveraging AD, \flinch{} overcomes the rigidity of hand-coded gradient pipelines. This architectural choice allows gradients to propagate from the pixel domain through the spherical-harmonic representation, making it straightforward to extend the model to the cosmological inference. We validated the framework on simulated, masked CMB temperature maps, demonstrating its capability to accurately reconstruct both maps and their angular power spectra, even in the presence of complex observational masks. By coupling \flinch{} with the differentiable CMB angular power-spectrum emulator \capse{}, we extended the analysis to perform cosmological parameter inference directly from the map, bypassing intermediate summary statistics. This approach retains the full statistical information content of the data. Our results show a clear advantage over the standard pseudo-\Cls{} methodology: when using the same simulated maps, \flinch{} achieves parameter constraints that are typically 20–40\% tighter. We also carried out an extensive performance comparison of three gradient-based samplers: HMC, NUTS, and MCLMC. MCLMC, in particular, showed exceptional scalability in high-dimensional parameter spaces, outperforming standard HMC by nearly three orders of magnitude in sampling efficiency for the highest resolution runs. When paired with a fully differentiable model, MCLMC enables computationally viable field-level inference at resolutions that would otherwise be extremely expensive. Overall, our findings position \flinch{} as a powerful, flexible, and efficient package for FLI. This combination of differentiable codes and advanced sampling strategies paves the way for applying this methodology to upcoming high-precision datasets from CMB and LSS surveys, potentially setting a new standard for cosmological data analysis.

Several extensions of this work provide promising avenues for future research. A natural next step is to extend \flinch{} to spin fields, following the same approach of~\cite{Loureiro_2023}, and interface it with a code such as \texttt{Blast.jl}, which can compute the full 3$\times$2pt statistics in an efficient and differentiable manner~\citep{Chiarenza:2024rgk}. This would enable an end-to-end, field-level cosmological analysis of surveys such as Euclid with the methods proposed here.

Another important direction is the use of cross-validation for systematics detection. As explored in \cite{Nguyen:2024spp}, such approaches can serve as a reliable and complementary alternative to more standard tools such as $\chi^2$ tests; in particular, Leave-One-Out and Leave-One-Group-Out cross-validation offer a more nuanced interpretation, with LOO posterior draws revealing the nature of the issues (e.g., under- or over-dispersed posterior predictive densities, model mis-specifications). This is particularly compelling for field-level inference: the tighter constraints it enables require correspondingly higher scrutiny, since undetected systematics can bias posterior estimates more severely than in summary-statistics analyses.

Finally, the methodology should be applied to real data, with a first ideal target being constraints on $f_{\mathrm{NL}}$~\citep{2023MNRAS.520.5746A, Euclid:2024ris}. This case is especially well-suited, as the signal resides predominantly on large scales where deviations from a Gaussian likelihood for summary statistics are most relevant~\citep{DESI:2023duv, fabbianquaia}. Moreover, even a low-resolution analysis with $n_{\mathrm{side}}=128$ could capture most of the available information, at a computational cost low enough to be performed within a couple of hours on a personal laptop.

\section*{Acknowledgments}

The authors are grateful to the \almanac{} team people (Alan Heavens, Andrew Jaffe, Elena Sellentin, Lorne Whiteway, Kutay Nazli, and Javier Lafaurie) for useful discussions.
MB acknowledges the support of the Natural Sciences and Engineering Research Council of Canada (NSERC), [funding
reference number RGPIN-2019-03908] from the Canadian Space Agency. MB acknowledges support from the INAF MiniGrant 2022. MB is grateful to David Alonso, Anton Baleato-Lizancos, Emanuele Castorina, Neal Dalal, Giulio Fabbian, Luigi Guzzo, Alex Krolewski, and Will Percival for useful discussions. AL acknowledges support from the Swedish National Space Agency (Rymdstyrelsen) under Career Grant Project Dnr 2024-00171 and from the research project grant `Understanding the Dynamic Universe' funded by the Knut and Alice Wallenberg Foundation under Dnr KAW 2018.0067.
This research was enabled in part by support provided by Compute Ontario (computeontario.ca) and the Digital Research Alliance of Canada (alliancecan.ca).
This research utilised the HPC facility supported by the Technical Division at the Department of Physics, Stockholm University.

\bibliographystyle{mnras}
\bibliography{biblio}

\begin{thebibliography}{}
\makeatletter
\relax
\def\mn@urlcharsother{\let\do\@makeother \do\$\do\&\do\#\do\^\do\_\do\%\do\~}
\def\mn@doi{\begingroup\mn@urlcharsother \@ifnextchar [ {\mn@doi@} {\mn@doi@[]}}
\def\mn@doi@[#1]#2{\def\@tempa{#1}\ifx\@tempa\@empty \href {http://dx.doi.org/#2} {doi:#2}\else \href {http://dx.doi.org/#2} {#1}\fi \endgroup}
\def\mn@eprint#1#2{\mn@eprint@#1:#2::\@nil}
\def\mn@eprint@arXiv#1{\href {http://arxiv.org/abs/#1} {{\tt arXiv:#1}}}
\def\mn@eprint@dblp#1{\href {http://dblp.uni-trier.de/rec/bibtex/#1.xml} {dblp:#1}}
\def\mn@eprint@#1:#2:#3:#4\@nil{\def\@tempa {#1}\def\@tempb {#2}\def\@tempc {#3}\ifx \@tempc \@empty \let \@tempc \@tempb \let \@tempb \@tempa \fi \ifx \@tempb \@empty \def\@tempb {arXiv}\fi \@ifundefined {mn@eprint@\@tempb}{\@tempb:\@tempc}{\expandafter \expandafter \csname mn@eprint@\@tempb\endcsname \expandafter{\@tempc}}}

\bibitem[\protect\citeauthoryear{Abbott et~al.}{Abbott et~al.}{2022}]{DES:2021wwk}
Abbott T. M.~C.,  et~al., 2022, \mn@doi [Phys. Rev. D] {10.1103/PhysRevD.105.023520}, 105, 023520

\bibitem[\protect\citeauthoryear{Abdul~Karim et~al.}{Abdul~Karim et~al.}{2025}]{DESI:2025fxa}
Abdul~Karim M.,  et~al., 2025, {Data Release 1 of the Dark Energy Spectroscopic Instrument} (\mn@eprint {arXiv} {2503.14745})

\bibitem[\protect\citeauthoryear{Ade et~al.}{Ade et~al.}{2019}]{SimonsObservatory:2018koc}
Ade P.,  et~al., 2019, \mn@doi [JCAP] {10.1088/1475-7516/2019/02/056}, 02, 056

\bibitem[\protect\citeauthoryear{Aghanim et~al.}{Aghanim et~al.}{2020}]{Planck:2018vyg}
Aghanim N.,  et~al., 2020, \mn@doi [Astron. Astrophys.] {10.1051/0004-6361/201833910}, 641, A6

\bibitem[\protect\citeauthoryear{Alam et~al.}{Alam et~al.}{2017}]{BOSS:2016wmc}
Alam S.,  et~al., 2017, \mn@doi [Mon. Not. Roy. Astron. Soc.] {10.1093/mnras/stx721}, 470, 2617

\bibitem[\protect\citeauthoryear{Alonso, Sanchez  \& Slosar}{Alonso et~al.}{2019}]{Alonso:2018jzx}
Alonso D.,  Sanchez J.,   Slosar A.,  2019, \mn@doi [Mon. Not. Roy. Astron. Soc.] {10.1093/mnras/stz093}, 484, 4127

\bibitem[\protect\citeauthoryear{Alsing, Heavens  \& Jaffe}{Alsing et~al.}{2017}]{Alsing:2016hkh}
Alsing J.,  Heavens A.~F.,   Jaffe A.~H.,  2017, \mn@doi [Mon. Not. Roy. Astron. Soc.] {10.1093/mnras/stw3161}, 466, 3272

\bibitem[\protect\citeauthoryear{{Alsing}, {Charnock}, {Feeney}  \& {Wandelt}}{{Alsing} et~al.}{2019}]{2019MNRAS.488.4440A}
{Alsing} J.,  {Charnock} T.,  {Feeney} S.,   {Wandelt} B.,  2019, \mn@doi [\mnras] {10.1093/mnras/stz1960}, \href {https://ui.adsabs.harvard.edu/abs/2019MNRAS.488.4440A} {488, 4440}

\bibitem[\protect\citeauthoryear{{Andrews}, {Jasche}, {Lavaux}  \& {Schmidt}}{{Andrews} et~al.}{2023}]{2023MNRAS.520.5746A}
{Andrews} A.,  {Jasche} J.,  {Lavaux} G.,   {Schmidt} F.,  2023, \mn@doi [\mnras] {10.1093/mnras/stad432}, \href {https://ui.adsabs.harvard.edu/abs/2023MNRAS.520.5746A} {520, 5746}

\bibitem[\protect\citeauthoryear{Andrews et~al.}{Andrews et~al.}{2024}]{Euclid:2024ris}
Andrews A.,  et~al., 2024, {Euclid: Field-level inference of primordial non-Gaussianity and cosmic initial conditions} (\mn@eprint {arXiv} {2412.11945})

\bibitem[\protect\citeauthoryear{Axen, Karrasch, Burton, Hauru, Yong  \& Qu}{Axen et~al.}{2025}]{seth_axen_2025_16818742}
Axen S.,  Karrasch D.,  Burton J.,  Hauru M.,  Yong P.,   Qu Q.,  2025, mlcolab/Pathfinder.jl: v0.9.21, \mn@doi{10.5281/zenodo.16818742}, \url {https://doi.org/10.5281/zenodo.16818742}

\bibitem[\protect\citeauthoryear{Balkenhol, Trendafilova, Benabed  \& Galli}{Balkenhol et~al.}{2024}]{Balkenhol:2024sbv}
Balkenhol L.,  Trendafilova C.,  Benabed K.,   Galli S.,  2024, \mn@doi [Astron. Astrophys.] {10.1051/0004-6361/202449432}, 686, A10

\bibitem[\protect\citeauthoryear{Bayer et~al.,}{Bayer et~al.}{2021}]{Bayer:2021iyb}
Bayer A.~E.,  et~al., 2021, \mn@doi [Astrophys. J.] {10.3847/1538-4357/ac0e91}, 919, 24

\bibitem[\protect\citeauthoryear{Bayer, Seljak  \& Modi}{Bayer et~al.}{2023}]{bayer2023}
Bayer A.~E.,  Seljak U.,   Modi C.,  2023, ICML 2023 Workshop on Machine Learning for Astrophysics

\bibitem[\protect\citeauthoryear{Bennett et~al.,}{Bennett et~al.}{2013}]{Bennett_2013}
Bennett C.~L.,  et~al., 2013, The Astrophysical Journal Supplement Series, 208, 20

\bibitem[\protect\citeauthoryear{Betancourt}{Betancourt}{2017}]{Betancourt2017}
Betancourt M.,  2017, A Conceptual Introduction to Hamiltonian Monte Carlo

\bibitem[\protect\citeauthoryear{Betancourt}{Betancourt}{2018}]{betancourt2018}
Betancourt M.,  2018, A Geometric Theory of Higher-Order Automatic Differentiation (\mn@eprint {arXiv} {1812.11592}), \url {https://arxiv.org/abs/1812.11592}

\bibitem[\protect\citeauthoryear{Bezanson, Karpinski, Shah  \& Edelman}{Bezanson et~al.}{2012}]{bezanson2012julia}
Bezanson J.,  Karpinski S.,  Shah V.~B.,   Edelman A.,  2012, arXiv preprint arXiv:1209.5145

\bibitem[\protect\citeauthoryear{Bianchi}{Bianchi}{2024}]{Bianchi2024}
Bianchi L.~A.,  2024, \mn@doi [Journal of Open Source Software] {10.21105/joss.06467}, 9, 6467

\bibitem[\protect\citeauthoryear{Bond}{Bond}{1995}]{Bond:1994aa}
Bond J.~R.,  1995, \mn@doi [Phys. Rev. Lett.] {10.1103/PhysRevLett.74.4369}, 74, 4369

\bibitem[\protect\citeauthoryear{Bonici et~al.}{Bonici et~al.}{2023}]{Euclid:2022hdx}
Bonici M.,  et~al., 2023, \mn@doi [Astron. Astrophys.] {10.1051/0004-6361/202244445}, 670, A47

\bibitem[\protect\citeauthoryear{Bonici, Bianchini  \& Ruiz-Zapatero}{Bonici et~al.}{2024a}]{Bonici_2024}
Bonici M.,  Bianchini F.,   Ruiz-Zapatero J.,  2024a, The Open Journal of Astrophysics, 7

\bibitem[\protect\citeauthoryear{Bonici, Biggio, Carbone  \& Guzzo}{Bonici et~al.}{2024b}]{Bonici:2022xlo}
Bonici M.,  Biggio L.,  Carbone C.,   Guzzo L.,  2024b, \mn@doi [Mon. Not. Roy. Astron. Soc.] {10.1093/mnras/stae1261}, 531, 4203

\bibitem[\protect\citeauthoryear{Bonici, D'Amico, Bel  \& Carbone}{Bonici et~al.}{2025}]{Bonici:2025ltp}
Bonici M.,  D'Amico G.,  Bel J.,   Carbone C.,  2025, \mn@doi [JCAP] {10.1088/1475-7516/2025/09/044}, 09, 044

\bibitem[\protect\citeauthoryear{Cagliari, Castorina, Bonici  \& Bianchi}{Cagliari et~al.}{2024}]{Cagliari:2023mkq}
Cagliari M.~S.,  Castorina E.,  Bonici M.,   Bianchi D.,  2024, \mn@doi [JCAP] {10.1088/1475-7516/2024/08/036}, 08, 036

\bibitem[\protect\citeauthoryear{Cagliari, Barberi-Squarotti, Pardede, Castorina  \& D'Amico}{Cagliari et~al.}{2025}]{Cagliari:2025rqe}
Cagliari M.~S.,  Barberi-Squarotti M.,  Pardede K.,  Castorina E.,   D'Amico G.,  2025, \mn@doi [JCAP] {10.1088/1475-7516/2025/07/043}, 07, 043

\bibitem[\protect\citeauthoryear{Campagne et~al.,}{Campagne et~al.}{2023}]{Campagne:2023ter}
Campagne J.-E.,  et~al., 2023, \mn@doi [Open J. Astrophys.] {10.21105/astro.2302.05163}, 6, 1

\bibitem[\protect\citeauthoryear{{Carron}}{{Carron}}{2013}]{carrongaussian}
{Carron} J.,  2013, \mn@doi [\aap] {10.1051/0004-6361/201220538}, \href {https://ui.adsabs.harvard.edu/abs/2013A&A...551A..88C} {551, A88}

\bibitem[\protect\citeauthoryear{Chaussidon et~al.}{Chaussidon et~al.}{2025}]{Chaussidon:2024qni}
Chaussidon E.,  et~al., 2025, \mn@doi [JCAP] {10.1088/1475-7516/2025/06/029}, 06, 029

\bibitem[\protect\citeauthoryear{Chen, Chen  \& Dvorkin}{Chen et~al.}{2025}]{Chen:2025wdy}
Chen S.-F.,  Chen K.-F.,   Dvorkin C.,  2025, {Field-level Reconstruction from Foreground-Contaminated 21-cm Maps} (\mn@eprint {arXiv} {2508.13265})

\bibitem[\protect\citeauthoryear{{Chiarenza}, {Bonici}, {Percival}  \& {White}}{{Chiarenza} et~al.}{2024}]{Chiarenza:2024rgk}
{Chiarenza} S.,  {Bonici} M.,  {Percival} W.,   {White} M.,  2024, \mn@doi [The Open Journal of Astrophysics] {10.33232/001c.127038}, \href {https://ui.adsabs.harvard.edu/abs/2024OJAp....7E.112C} {7, 112}

\bibitem[\protect\citeauthoryear{Contarini et~al.}{Contarini et~al.}{2022}]{Euclid:2022qtk}
Contarini S.,  et~al., 2022, \mn@doi [Astron. Astrophys.] {10.1051/0004-6361/202244095}, 667, A162

\bibitem[\protect\citeauthoryear{Cranmer, Brehmer  \& Louppe}{Cranmer et~al.}{2020}]{doi:10.1073/pnas.1912789117}
Cranmer K.,  Brehmer J.,   Louppe G.,  2020, \mn@doi [Proceedings of the National Academy of Sciences] {10.1073/pnas.1912789117}, 117, 30055

\bibitem[\protect\citeauthoryear{D'Amico, Donath, Lewandowski, Senatore  \& Zhang}{D'Amico et~al.}{2024}]{DAmico:2022osl}
D'Amico G.,  Donath Y.,  Lewandowski M.,  Senatore L.,   Zhang P.,  2024, \mn@doi [JCAP] {10.1088/1475-7516/2024/05/059}, 05, 059

\bibitem[\protect\citeauthoryear{Dalal, Dore, Huterer  \& Shirokov}{Dalal et~al.}{2008}]{Dalal:2007cu}
Dalal N.,  Dore O.,  Huterer D.,   Shirokov A.,  2008, \mn@doi [Phys. Rev. D] {10.1103/PhysRevD.77.123514}, 77, 123514

\bibitem[\protect\citeauthoryear{Duane, Kennedy, Pendleton  \& Roweth}{Duane et~al.}{1987}]{DUANE1987216}
Duane S.,  Kennedy A.,  Pendleton B.~J.,   Roweth D.,  1987, \mn@doi [Physics Letters B] {https://doi.org/10.1016/0370-2693(87)91197-X}, 195, 216

\bibitem[\protect\citeauthoryear{Efstathiou}{Efstathiou}{2004}]{Efstathiou:2003dj}
Efstathiou G.,  2004, \mn@doi [Mon. Not. Roy. Astron. Soc.] {10.1111/j.1365-2966.2004.07530.x}, 349, 603

\bibitem[\protect\citeauthoryear{{Fabbian}, {Alonso}, {Storey-Fisher}  \& {Cornish}}{{Fabbian} et~al.}{2025}]{fabbianquaia}
{Fabbian} G.,  {Alonso} D.,  {Storey-Fisher} K.,   {Cornish} T.,  2025, \mn@doi [arXiv e-prints] {10.48550/arXiv.2504.20992}, p. arXiv:2504.20992

\bibitem[\protect\citeauthoryear{Fjelde et~al.,}{Fjelde et~al.}{2025}]{10.1145/3711897}
Fjelde T.~E.,  et~al., 2025, \mn@doi [ACM Trans. Probab. Mach. Learn.] {10.1145/3711897}

\bibitem[\protect\citeauthoryear{Galloway et~al.}{Galloway et~al.}{2023}]{Galloway:2022rhr}
Galloway M.,  et~al., 2023, \mn@doi [Astron. Astrophys.] {10.1051/0004-6361/202243138}, 675, A8

\bibitem[\protect\citeauthoryear{García-García, Alonso  \& Bellini}{García-García et~al.}{2019}]{Garc_a_Garc_a_2019}
García-García C.,  Alonso D.,   Bellini E.,  2019, \mn@doi [Journal of Cosmology and Astroparticle Physics] {10.1088/1475-7516/2019/11/043}, 2019, 043–043

\bibitem[\protect\citeauthoryear{Gatti et~al.}{Gatti et~al.}{2025}]{DES:2024jgw}
Gatti M.,  et~al., 2025, \mn@doi [Phys. Rev. D] {10.1103/PhysRevD.111.063504}, 111, 063504

\bibitem[\protect\citeauthoryear{Gelman \& Rubin}{Gelman \& Rubin}{1992}]{gelman_rubin}
Gelman A.,  Rubin D.~B.,  1992, Statistical Science, 7, 457

\bibitem[\protect\citeauthoryear{Ghigna et~al.}{Ghigna et~al.}{2024}]{LiteBIRD:2024wix}
Ghigna T.,  et~al., 2024, in {SPIE Astronomical Telescopes + Instrumentation 2024}.  (\mn@eprint {arXiv} {2406.02724})

\bibitem[\protect\citeauthoryear{Gorski}{Gorski}{1994}]{Gorski:1994ye}
Gorski K.~M.,  1994, \mn@doi [Astrophys. J. Lett.] {10.1086/187444}, 430, L85

\bibitem[\protect\citeauthoryear{Gorski, Hivon, Banday, Wandelt, Hansen, Reinecke  \& Bartelmann}{Gorski et~al.}{2005}]{Gorski_2005}
Gorski K.~M.,  Hivon E.,  Banday A.~J.,  Wandelt B.~D.,  Hansen F.~K.,  Reinecke M.,   Bartelmann M.,  2005, The Astrophysical Journal, 622, 759–771

\bibitem[\protect\citeauthoryear{Griewank \& Walther}{Griewank \& Walther}{2008}]{griewank2008evaluating}
Griewank A.,  Walther A.,  2008, Evaluating Derivatives: Principles and Techniques of Algorithmic Differentiation, Second Edition.
Other Titles in Applied Mathematics, Society for Industrial and Applied Mathematics, \url {https://books.google.ca/books?id=qMLUIsgCwvUC}

\bibitem[\protect\citeauthoryear{Hahn et~al.,}{Hahn et~al.}{2023}]{Hahn:2022wgo}
Hahn C.,  et~al., 2023, \mn@doi [JCAP] {10.1088/1475-7516/2023/04/010}, 04, 010

\bibitem[\protect\citeauthoryear{Hahn, List  \& Porqueres}{Hahn et~al.}{2024}]{Hahn:2023nvb}
Hahn O.,  List F.,   Porqueres N.,  2024, \mn@doi [JCAP] {10.1088/1475-7516/2024/06/063}, 06, 063

\bibitem[\protect\citeauthoryear{Hamaus et~al.}{Hamaus et~al.}{2022}]{Euclid:2021xmh}
Hamaus N.,  et~al., 2022, \mn@doi [Astron. Astrophys.] {10.1051/0004-6361/202142073}, 658, A20

\bibitem[\protect\citeauthoryear{Hamilton}{Hamilton}{2008a}]{Hamilton:2005kz}
Hamilton A. J.~S.,  2008a, Lect. Notes Phys., 665, 415

\bibitem[\protect\citeauthoryear{Hamilton}{Hamilton}{2008b}]{Hamilton:2005ma}
Hamilton A. J.~S.,  2008b, Lect. Notes Phys., 665, 433

\bibitem[\protect\citeauthoryear{Hamimeche \& Lewis}{Hamimeche \& Lewis}{2008}]{Hamimeche:2008ai}
Hamimeche S.,  Lewis A.,  2008, \mn@doi [Phys. Rev. D] {10.1103/PhysRevD.77.103013}, 77, 103013

\bibitem[\protect\citeauthoryear{Hinshaw et~al.,}{Hinshaw et~al.}{2013}]{Hinshaw_2013}
Hinshaw G.,  et~al., 2013, The Astrophysical Journal Supplement Series, 208, 19

\bibitem[\protect\citeauthoryear{Hoffman \& Gelman}{Hoffman \& Gelman}{2011}]{HoffmanGelman2011}
Hoffman M.~D.,  Gelman A.,  2011, Journal of Machine Learning Research, 15, 1593

\bibitem[\protect\citeauthoryear{Hu \& White}{Hu \& White}{1997}]{Hu:1997hp}
Hu W.,  White M.~J.,  1997, \mn@doi [Phys. Rev. D] {10.1103/PhysRevD.56.596}, 56, 596

\bibitem[\protect\citeauthoryear{Innes}{Innes}{2019}]{innes2019}
Innes M.,  2019, Don't Unroll Adjoint: Differentiating SSA-Form Programs (\mn@eprint {arXiv} {1810.07951}), \url {https://arxiv.org/abs/1810.07951}

\bibitem[\protect\citeauthoryear{Ivezi{\'{c}} et~al.,}{Ivezi{\'{c}} et~al.}{2019}]{Ivezic2019LSST:Products}
Ivezi{\'{c}} {\v{Z}}.,  et~al., 2019, \mn@doi [The Astrophysical Journal] {10.3847/1538-4357/ab042c}, 873, 111

\bibitem[\protect\citeauthoryear{{Jasche} \& {Wandelt}}{{Jasche} \& {Wandelt}}{2013}]{2013MNRAS.432..894J}
{Jasche} J.,  {Wandelt} B.~D.,  2013, \mn@doi [\mnras] {10.1093/mnras/stt449}, \href {https://ui.adsabs.harvard.edu/abs/2013MNRAS.432..894J} {432, 894}

\bibitem[\protect\citeauthoryear{Jeffrey et~al.}{Jeffrey et~al.}{2024}]{DES:2024xij}
Jeffrey N.,  et~al., 2024, \mn@doi [Mon. Not. Roy. Astron. Soc.] {10.1093/mnras/stae2629}, 536, 1303

\bibitem[\protect\citeauthoryear{{Jewell}, {Eriksen}, {Wandelt}, {O'Dwyer}, {Huey}  \& {G{\'o}rski}}{{Jewell} et~al.}{2009}]{2009ApJ...697..258J}
{Jewell} J.~B.,  {Eriksen} H.~K.,  {Wandelt} B.~D.,  {O'Dwyer} I.~J.,  {Huey} G.,   {G{\'o}rski} K.~M.,  2009, \mn@doi [\apj] {10.1088/0004-637X/697/1/258}, \href {https://ui.adsabs.harvard.edu/abs/2009ApJ...697..258J} {697, 258}

\bibitem[\protect\citeauthoryear{Krause et~al.}{Krause et~al.}{2025}]{Beyond-2pt:2024mqz}
Krause E.,  et~al., 2025, \mn@doi [Astrophys. J.] {10.3847/1538-4357/ad781d}, 990, 99

\bibitem[\protect\citeauthoryear{Krolewski et~al.}{Krolewski et~al.}{2024}]{DESI:2023duv}
Krolewski A.,  et~al., 2024, \mn@doi [JCAP] {10.1088/1475-7516/2024/03/021}, 03, 021

\bibitem[\protect\citeauthoryear{Kubo}{Kubo}{1966}]{Kubo:1966fyg}
Kubo R.,  1966, \mn@doi [Rept. Prog. Phys.] {10.1088/0034-4885/29/1/306}, 29, 255

\bibitem[\protect\citeauthoryear{{Leclercq} \& {Heavens}}{{Leclercq} \& {Heavens}}{2021}]{2021MNRAS.506L..85L}
{Leclercq} F.,  {Heavens} A.,  2021, \mn@doi [\mnras] {10.1093/mnrasl/slab081}, \href {https://ui.adsabs.harvard.edu/abs/2021MNRAS.506L..85L} {506, L85}

\bibitem[\protect\citeauthoryear{Leistedt, Peiris, Mortlock, Benoit-L{\'e}vy  \& Pontzen}{Leistedt et~al.}{2013}]{Leistedt:2013gfa}
Leistedt B.,  Peiris H.~V.,  Mortlock D.~J.,  Benoit-L{\'e}vy A.,   Pontzen A.,  2013, \mn@doi [Mon. Not. Roy. Astron. Soc.] {10.1093/mnras/stt1359}, 435, 1857

\bibitem[\protect\citeauthoryear{Li et~al.,}{Li et~al.}{2022}]{Li:2022qlf}
Li Y.,  et~al., 2022, {pmwd: A Differentiable Cosmological Particle-Mesh $N$-body Library} (\mn@eprint {arXiv} {2211.09958})

\bibitem[\protect\citeauthoryear{Liu \& Nocedal}{Liu \& Nocedal}{1989}]{LiuNocedal1989}
Liu D.~C.,  Nocedal J.,  1989, \mn@doi [Mathematical Programming] {10.1007/BF01589116}, 45, 503

\bibitem[\protect\citeauthoryear{Loureiro, Whiteaway, Sellentin, Lafaurie, Jaffe  \& Heavens}{Loureiro et~al.}{2023}]{Loureiro_2023}
Loureiro A.,  Whiteaway L.,  Sellentin E.,  Lafaurie J.~S.,  Jaffe A.~H.,   Heavens A.~F.,  2023, The Open Journal of Astrophysics, 6

\bibitem[\protect\citeauthoryear{Makinen, Charnock, Alsing  \& Wandelt}{Makinen et~al.}{2021}]{Makinen:2021nly}
Makinen T.~L.,  Charnock T.,  Alsing J.,   Wandelt B.~D.,  2021, \mn@doi [JCAP] {10.1088/1475-7516/2021/11/049}, 11, 049

\bibitem[\protect\citeauthoryear{Maraio, Hall  \& Taylor}{Maraio et~al.}{2023}]{Maraio:2022ywi}
Maraio A.,  Hall A.,   Taylor A.,  2023, \mn@doi [Mon. Not. Roy. Astron. Soc.] {10.1093/mnras/stad426}, 520, 4836

\bibitem[\protect\citeauthoryear{Mellier et~al.}{Mellier et~al.}{2025}]{Euclid:2024yrr}
Mellier Y.,  et~al., 2025, \mn@doi [Astron. Astrophys.] {10.1051/0004-6361/202450810}, 697, A1

\bibitem[\protect\citeauthoryear{Meng \& Rubin}{Meng \& Rubin}{1991}]{meng_rubin}
Meng X.-L.,  Rubin D.~B.,  1991, Journal of the American Statistical Association, 86, 899

\bibitem[\protect\citeauthoryear{Millea, Anderes  \& Wandelt}{Millea et~al.}{2019}]{Millea:2017fyd}
Millea M.,  Anderes E.,   Wandelt B.~D.,  2019, \mn@doi [Phys. Rev. D] {10.1103/PhysRevD.100.023509}, 100, 023509

\bibitem[\protect\citeauthoryear{Nazli et~al.}{Nazli et~al.}{2025}]{Nazli_inprep}
Nazli K.,  et~al., 2025, in preparation

\bibitem[\protect\citeauthoryear{Neal}{Neal}{2003}]{feba456f-8373-3f4e-b93a-903e0caf3a0f}
Neal R.~M.,  2003, The Annals of Statistics, 31, 705

\bibitem[\protect\citeauthoryear{Nesterov}{Nesterov}{2009}]{Nesterov2009}
Nesterov Y.,  2009, \mn@doi [Mathematical Programming] {10.1007/s10107-007-0149-x}, 120, 221

\bibitem[\protect\citeauthoryear{Nguyen, Bonici, McGee  \& Percival}{Nguyen et~al.}{2025}]{Nguyen:2024spp}
Nguyen A. B.~H.,  Bonici M.,  McGee G.,   Percival W.~J.,  2025, \mn@doi [JCAP] {10.1088/1475-7516/2025/01/008}, 01, 008

\bibitem[\protect\citeauthoryear{Oehl \& Tr{\" o}ster}{Oehl \& Tr{\" o}ster}{2025}]{Oehl:2024gbm}
Oehl V.,  Tr{\" o}ster T.,  2025, \mn@doi [The Open Journal of Astrophysics] {10.33232/001c.144235}, 8

\bibitem[\protect\citeauthoryear{Paillas et~al.}{Paillas et~al.}{2024}]{Paillas:2023cpk}
Paillas E.,  et~al., 2024, \mn@doi [Mon. Not. Roy. Astron. Soc.] {10.1093/mnras/stae1118}, 531, 898

\bibitem[\protect\citeauthoryear{Papaspiliopoulos, Roberts  \& Sköld}{Papaspiliopoulos et~al.}{2007}]{Papaspiliopoulos}
Papaspiliopoulos O.,  Roberts G.~O.,   Sköld M.,  2007, Statistical Science, 22, 59

\bibitem[\protect\citeauthoryear{Paradiso et~al.}{Paradiso et~al.}{2023}]{Paradiso:2022fky}
Paradiso S.,  et~al., 2023, \mn@doi [Astron. Astrophys.] {10.1051/0004-6361/202244060}, 675, A12

\bibitem[\protect\citeauthoryear{{Peebles}}{{Peebles}}{1973}]{1973ApJ...185..413P}
{Peebles} P.~J.~E.,  1973, \mn@doi [\apj] {10.1086/152431}, \href {https://ui.adsabs.harvard.edu/abs/1973ApJ...185..413P} {185, 413}

\bibitem[\protect\citeauthoryear{Philcox}{Philcox}{2021a}]{Philcox:2020vbm}
Philcox O. H.~E.,  2021a, \mn@doi [Phys. Rev. D] {10.1103/PhysRevD.103.103504}, 103, 103504

\bibitem[\protect\citeauthoryear{Philcox}{Philcox}{2021b}]{Philcox:2021ukg}
Philcox O. H.~E.,  2021b, \mn@doi [Phys. Rev. D] {10.1103/PhysRevD.104.123529}, 104, 123529

\bibitem[\protect\citeauthoryear{Philcox \& Ivanov}{Philcox \& Ivanov}{2022}]{Philcox:2021kcw}
Philcox O. H.~E.,  Ivanov M.~M.,  2022, \mn@doi [Phys. Rev. D] {10.1103/PhysRevD.105.043517}, 105, 043517

\bibitem[\protect\citeauthoryear{Piras \& Spurio~Mancini}{Piras \& Spurio~Mancini}{2023}]{Piras2023CosmoPower}
Piras D.,  Spurio~Mancini A.,  2023, \mn@doi [The Open Journal of Astrophysics] {10.21105/astro.2305.06347}, 6

\bibitem[\protect\citeauthoryear{{Pr{\'e}zeau} \& {Reinecke}}{{Pr{\'e}zeau} \& {Reinecke}}{2010}]{2010ApJS..190..267P}
{Pr{\'e}zeau} G.,  {Reinecke} M.,  2010, \mn@doi [\apjs] {10.1088/0067-0049/190/2/267}, 190, 267

\bibitem[\protect\citeauthoryear{{Price} \& {McEwen}}{{Price} \& {McEwen}}{2024}]{2024JCoPh.51013109P}
{Price} M.~A.,  {McEwen} J.~D.,  2024, \mn@doi [Journal of Computational Physics] {10.1016/j.jcp.2024.113109}, \href {https://ui.adsabs.harvard.edu/abs/2024JCoPh.51013109P} {510, 113109}

\bibitem[\protect\citeauthoryear{Reinecke \& Seljebotn}{Reinecke \& Seljebotn}{2013}]{Reinecke_2013}
Reinecke M.,  Seljebotn D.~S.,  2013, Astronomy \& Astrophysics, 554, A112

\bibitem[\protect\citeauthoryear{Robnik \& Seljak}{Robnik \& Seljak}{2023}]{Robnik:2023pgt}
Robnik J.,  Seljak U.,  2023, {Fluctuation without dissipation: Microcanonical Langevin Monte Carlo} (\mn@eprint {arXiv} {2303.18221})

\bibitem[\protect\citeauthoryear{Robnik, De~Luca, Silverstein  \& Seljak}{Robnik et~al.}{2023}]{Robnik2022}
Robnik J.,  De~Luca G.~B.,  Silverstein E.,   Seljak U.,  2023, Journal of Machine Learning Research, 24, 1

\bibitem[\protect\citeauthoryear{{Robnik}, {Cohn-Gordon}  \& {Seljak}}{{Robnik} et~al.}{2025}]{2025arXiv250301707R}
{Robnik} J.,  {Cohn-Gordon} R.,   {Seljak} U.,  2025, \mn@doi [arXiv e-prints] {10.48550/arXiv.2503.01707}, \href {https://ui.adsabs.harvard.edu/abs/2025arXiv250301707R} {p. arXiv:2503.01707}

\bibitem[\protect\citeauthoryear{Ruiz-Zapatero, Alonso, Garc{\' i}a-Garc{\' i}a, Nicola, Mootoovaloo, Sullivan, Bonici  \& G.~Ferreira}{Ruiz-Zapatero et~al.}{2024}]{Ruiz2024LimberJack}
Ruiz-Zapatero J.,  Alonso D.,  Garc{\' i}a-Garc{\' i}a C.,  Nicola A.,  Mootoovaloo A.,  Sullivan J.~M.,  Bonici M.,   G.~Ferreira P.,  2024, \mn@doi [The Open Journal of Astrophysics] {10.21105/astro.2310.08306}, 7

\bibitem[\protect\citeauthoryear{Schmidt}{Schmidt}{2021}]{Schmidt:2020ovm}
Schmidt F.,  2021, \mn@doi [JCAP] {10.1088/1475-7516/2021/04/033}, 04, 033

\bibitem[\protect\citeauthoryear{Seljak, Aslanyan, Feng  \& Modi}{Seljak et~al.}{2017}]{Seljak:2017rmr}
Seljak U.,  Aslanyan G.,  Feng Y.,   Modi C.,  2017, \mn@doi [JCAP] {10.1088/1475-7516/2017/12/009}, 12, 009

\bibitem[\protect\citeauthoryear{Sellentin \& Heavens}{Sellentin \& Heavens}{2016}]{Sellentin:2015waz}
Sellentin E.,  Heavens A.~F.,  2016, \mn@doi [Mon. Not. Roy. Astron. Soc.] {10.1093/mnrasl/slv190}, 456, L132

\bibitem[\protect\citeauthoryear{Sellentin, Jaffe  \& Heavens}{Sellentin et~al.}{2017}]{Sellentin:2017aii}
Sellentin E.,  Jaffe A.~H.,   Heavens A.~F.,  2017, {On the use of the Edgeworth expansion in cosmology I: how to foresee and evade its pitfalls} (\mn@eprint {arXiv} {1709.03452})

\bibitem[\protect\citeauthoryear{Sellentin, Loureiro, Whiteway, Lafaurie, Balan, Olamaie, Jaffe  \& Heavans}{Sellentin et~al.}{2023}]{Sellentin_2023}
Sellentin E.,  Loureiro A.,  Whiteway L.,  Lafaurie J.~S.,  Balan S.~T.,  Olamaie M.,  Jaffe A.~H.,   Heavans A.~F.,  2023, The Open Journal of Astrophysics, 6

\bibitem[\protect\citeauthoryear{Simon-Onfroy, Lanusse  \& de Mattia}{Simon-Onfroy et~al.}{2025}]{Simon-Onfroy:2025ziw}
Simon-Onfroy H.,  Lanusse F.,   de Mattia A.,  2025, Benchmarking field-level cosmological inference from galaxy redshift surveys

\bibitem[\protect\citeauthoryear{Sletmoen}{Sletmoen}{2025}]{Sletmoen:2025fro}
Sletmoen H.,  2025, {SymBoltz.jl: a symbolic-numeric, approximation-free and differentiable linear Einstein-Boltzmann solver} (\mn@eprint {arXiv} {2509.24740})

\bibitem[\protect\citeauthoryear{Spurio~Mancini, Lin  \& McEwen}{Spurio~Mancini et~al.}{2024}]{SpurioMancini:2024qic}
Spurio~Mancini A.,  Lin K.,   McEwen J.~D.,  2024, {Field-level cosmological model selection: field-level simulation-based inference for Stage IV cosmic shear can distinguish dynamical dark energy} (\mn@eprint {arXiv} {2410.10616})

\bibitem[\protect\citeauthoryear{Sullivan, Hergt  \& Scott}{Sullivan et~al.}{2024}]{sullivan2024}
Sullivan R.~M.,  Hergt L.~T.,   Scott D.,  2024, Methods for CMB map analysis (\mn@eprint {arXiv} {2410.12951}), \url {https://arxiv.org/abs/2410.12951}

\bibitem[\protect\citeauthoryear{Sunseri, Bayer  \& Liu}{Sunseri et~al.}{2025}]{Sunseri:2025jem}
Sunseri J.,  Bayer A.~E.,   Liu J.,  2025, \mn@doi [Phys. Rev. D] {10.1103/grx3-hj7w}, 112, 063516

\bibitem[\protect\citeauthoryear{{Taylor}, {Ashdown}  \& {Hobson}}{{Taylor} et~al.}{2008}]{2008MNRAS.389.1284T}
{Taylor} J.~F.,  {Ashdown} M.~A.~J.,   {Hobson} M.~P.,  2008, \mn@doi [\mnras] {10.1111/j.1365-2966.2008.13630.x}, \href {https://ui.adsabs.harvard.edu/abs/2008MNRAS.389.1284T} {389, 1284}

\bibitem[\protect\citeauthoryear{Tegmark}{Tegmark}{1997}]{Tegmark:1996qt}
Tegmark M.,  1997, \mn@doi [Phys. Rev. D] {10.1103/PhysRevD.55.5895}, 55, 5895

\bibitem[\protect\citeauthoryear{Tegmark \& de Oliveira-Costa}{Tegmark \& de~Oliveira-Costa}{2001}]{Tegmark:2001zv}
Tegmark M.,  de Oliveira-Costa A.,  2001, \mn@doi [Phys. Rev. D] {10.1103/PhysRevD.64.063001}, 64, 063001

\bibitem[\protect\citeauthoryear{Valogiannis, Villaescusa-Navarro  \& Baldi}{Valogiannis et~al.}{2024}]{Valogiannis:2024rvt}
Valogiannis G.,  Villaescusa-Navarro F.,   Baldi M.,  2024, \mn@doi [JCAP] {10.1088/1475-7516/2024/11/061}, 11, 061

\bibitem[\protect\citeauthoryear{Vehtari, Simpson, Gelman, Yao  \& Gabry}{Vehtari et~al.}{2024}]{JMLR:v25:19-556}
Vehtari A.,  Simpson D.,  Gelman A.,  Yao Y.,   Gabry J.,  2024, Journal of Machine Learning Research, 25, 1

\bibitem[\protect\citeauthoryear{Wandelt \& Gorski}{Wandelt \& Gorski}{2001}]{Wandelt:2001gp}
Wandelt B.~D.,  Gorski K.~M.,  2001, \mn@doi [Phys. Rev. D] {10.1103/PhysRevD.63.123002}, 63, 123002

\bibitem[\protect\citeauthoryear{{Wandelt}, {Larson}  \& {Lakshminarayanan}}{{Wandelt} et~al.}{2004}]{2004PhRvD..70h3511W}
{Wandelt} B.~D.,  {Larson} D.~L.,   {Lakshminarayanan} A.,  2004, \mn@doi [\prd] {10.1103/PhysRevD.70.083511}, \href {https://ui.adsabs.harvard.edu/abs/2004PhRvD..70h3511W} {70, 083511}

\bibitem[\protect\citeauthoryear{Wright et~al.}{Wright et~al.}{2025}]{Wright:2025xka}
Wright A.~H.,  et~al., 2025, {KiDS-Legacy: Cosmological constraints from cosmic shear with the complete Kilo-Degree Survey} (\mn@eprint {arXiv} {2503.19441})

\bibitem[\protect\citeauthoryear{Zhang, Carpenter, Gelman  \& Vehtari}{Zhang et~al.}{2022}]{zhang_2022}
Zhang L.,  Carpenter B.,  Gelman A.,   Vehtari A.,  2022, Journal of Machine Learning Research, 23, 1

\bibitem[\protect\citeauthoryear{von Wietersheim-Kramsta, Lin, Tessore, Joachimi, Loureiro, Reischke  \& Wright}{von Wietersheim-Kramsta et~al.}{2025}]{vonWietersheim-Kramsta:2024cks}
von Wietersheim-Kramsta M.,  Lin K.,  Tessore N.,  Joachimi B.,  Loureiro A.,  Reischke R.,   Wright A.~H.,  2025, \mn@doi [Astron. Astrophys.] {10.1051/0004-6361/202450487}, 694, A223

\makeatother
\end{thebibliography}

\appendix

\section{Convergence and performance}
\label{app:convergence_perf}

\subsection{Chains' convergence}
Monitoring the convergence of a Monte Carlo chain in a parameter space of such high dimensionality (in this test, exceeding approximately one million of parameters) is a task that requires care.
As a first step, we verified that the Gelman–Rubin factor (\(\hat{R}\))~\citep{gelman_rubin}, which quantifies the ratio between the within-chain variance and the between-chains variance, deviated from unity within the thresholds typically considered indicative of good convergence, namely, within 5–10\%.
As shown in Fig.~\ref{fig:Cls_GelmanRubin}, for the \Cls, which are central to the analysis, both NUTS and MCLMC achieve \(\hat{R}\) values below 1.01 for the majority of multipoles. HMC, on the other hand, reaches convergence with \(\hat{R}\) values below 1.05, which reflects the need for longer chains to ensure better mixing.
When examining the \(\hat{R}\) factor for the reconstructed modes \alm, we confirm that NUTS and MCLMC chains converge within 1\%, whereas HMC convergence for these parameters degrades to approximately 10\%.
\begin{figure}[!ht]
\centering
\includegraphics[width=.5\textwidth]{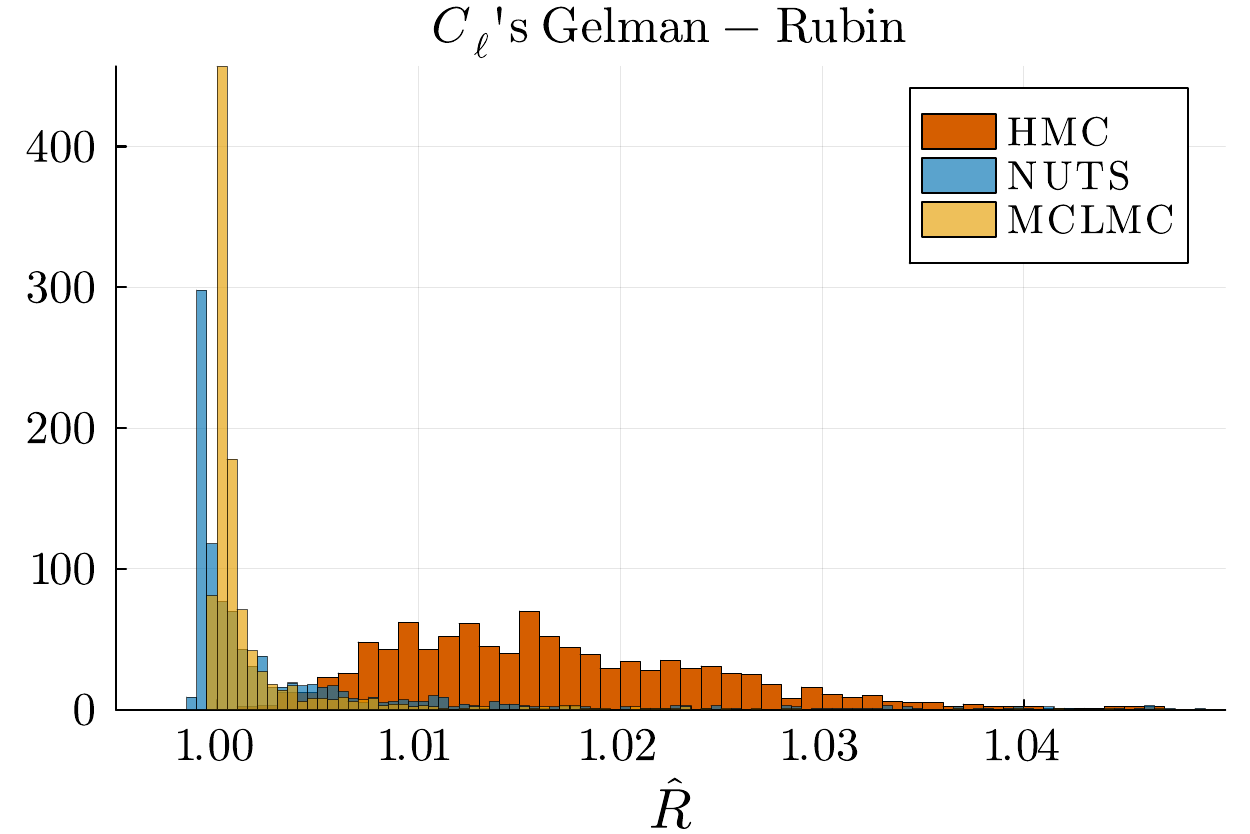}
\caption{Histogram of Gelman-Rubin statistics for the reconstructed $C_\ell$'s. Most $C_\ell$'s from NUTS and MCLMC achieveve $\hat{R}<1.01$, indicating excellent convergence, whereas HMC concentrates below $\hat{R}<1.05$, reflecting comparatively lower convergence.}
\label{fig:Cls_GelmanRubin}
\end{figure}
For both NUTS and HMC, we also evaluated the Bayesian Fraction of Missing Information (BFMI)~\citep{meng_rubin}, a diagnostic tool designed to assess how efficiently the sampler is exploring the energy distribution of the target posterior. The BFMI compares the variance of the energy transitions observed along the Markov chain to the marginal variance of the energy under the target distribution. A low BFMI, typically below 0.3, indicates poor exploration of the posterior energy, often resulting in biased or inefficient sampling, while values greater than 0.7 signal good mixing.
This diagnostic is typically interpreted in conjunction with the energy histogram plots. These compare the distribution of the total energy at each step \(E_k = H_k-\bar{H}\) with the distribution of energy transitions \(\Delta E_k=H_{k+1}-H_k\). If \(\Delta E_k\) varies much less than \(E_k\), the sampler is not moving efficiently through the energy space, resulting in a low BFMI. In contrast, when the two histograms have similar width, the sampler is making efficient energy transitions, leading to a higher BFMI and better mixing.
In our analysis, NUTS achieves BFMI values consistently above 0.9, with overlapping and similarly shaped energy histograms, indicating excellent exploration. HMC instead yields BFMI values around 0.5, and its energy transition histogram is noticeably narrower than that of the total energy, consistent with less effective mixing and the convergence issues discussed previously.
We do not compute the BFMI diagnostic for MCLMC, as its dynamics is governed by a microcanonical Hamiltonian system, where the energy is by construction nearly conserved. Consequently, energy transitions are null, and the assumptions underlying the BFMI statistic are no longer valid. In this context, BFMI does not provide meaningful information about sampling efficiency or convergence. As a final diagnostic, we evaluate the integrated autocorrelation time of the chains, which quantifies how many steps are needed before samples become effectively uncorrelated. This metric is crucial for assessing the statistical efficiency of a sampler, as it directly impacts the effective number of independent samples obtained from a finite chain.
We compute the autocorrelation time for each parameter. In particular, for the power spectrum multipoles \Cls, we find integrated autocorrelation times of less than 10 steps for NUTS, 20 for MCLMC, and approximately 100 for HMC. The \alm{} parameters typically exhibit autocorrelation times that are an order of magnitude larger across all samplers. Finally, for MCLMC we also verified the absence of significant biases in the chains for the various \Cls{} parameters. Using the mean and standard deviation obtained from NUTS as a reference target, since NUTS is by design unbiased and showed better convergence than HMC, we plot the relative bias between the cumulative mean of each sampler and the target values, normalized by the target standard deviation, as a function of the number of gradient evaluations.
The curves are computed separately for each chain in order to display an average trend. This approach allows us to identify the bias floor below which MCLMC does not improve further, while also providing insight into the relative efficiency of the samplers by showing how many fewer gradient evaluations MCLMC requires to reach stability.
In Fig.~\ref{fig:mean_bias_gradevals} we present this diagnostic plot for a selection of multipole values. 
On average, the magnitude of the final bias in the mean introduced by MCLMC remains smaller than the standard Monte Carlo error of NUTS, making it entirely negligible and confirming the reliability of the sampler.

\begin{figure}[htbp]
    \centering
    \begin{subfigure}{0.4\textwidth}
        \includegraphics[width=\linewidth]{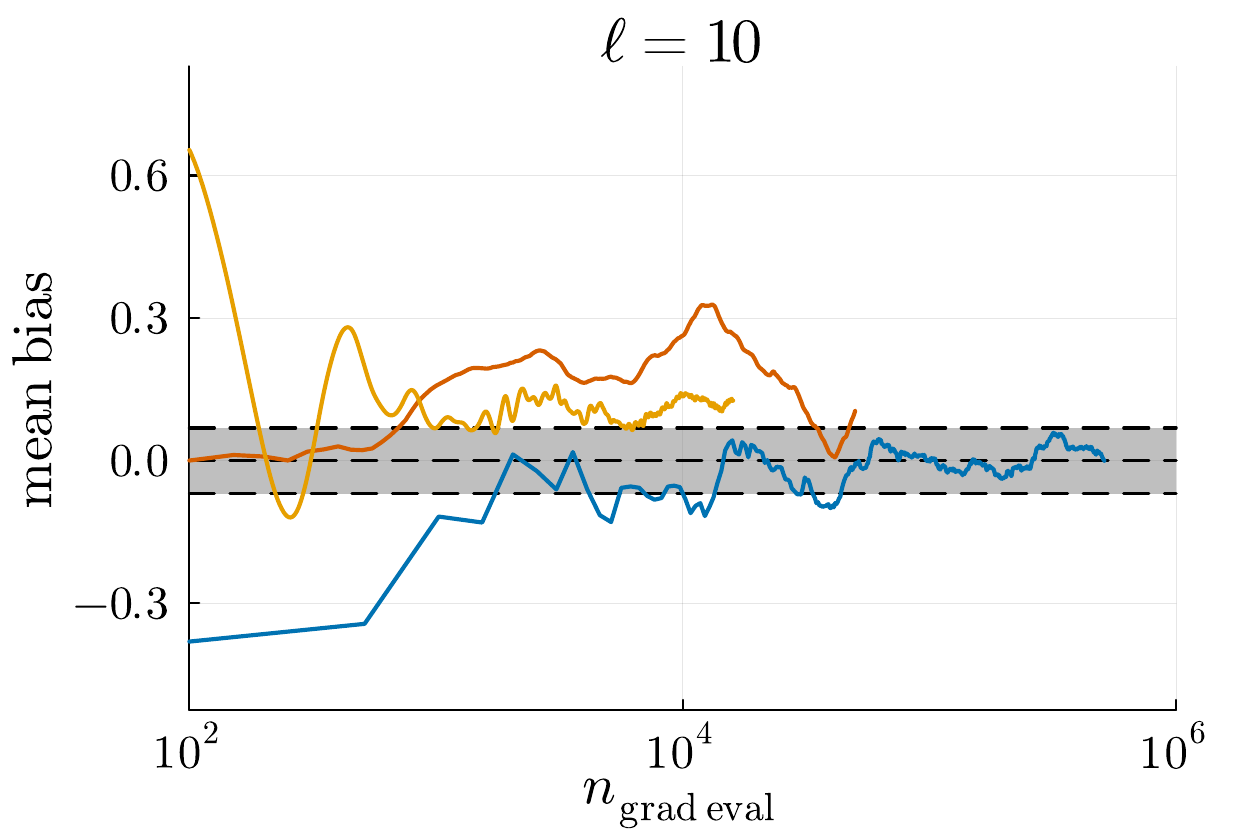}
    \end{subfigure}
    \begin{subfigure}{0.4\textwidth}
        \includegraphics[width=\linewidth]{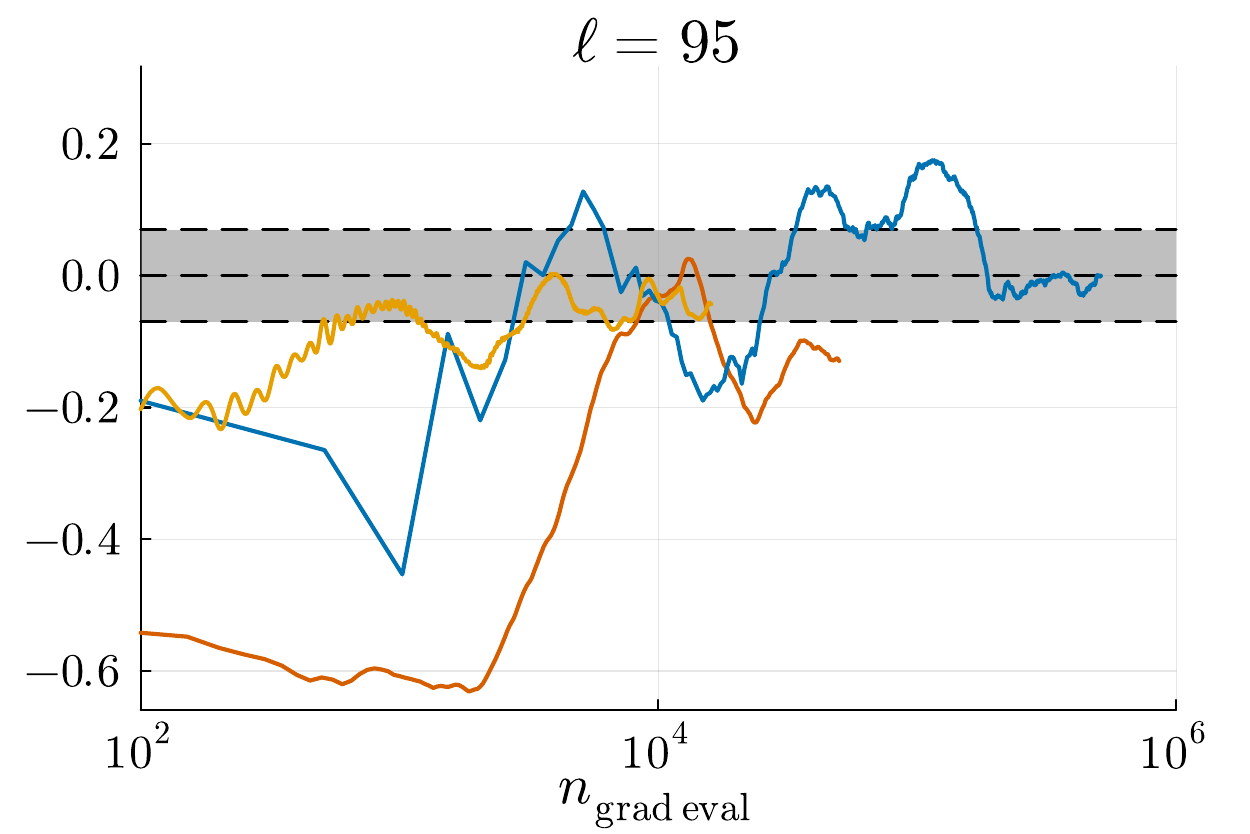}
    \end{subfigure}
    \begin{subfigure}{0.4\textwidth}
        \includegraphics[width=\linewidth]{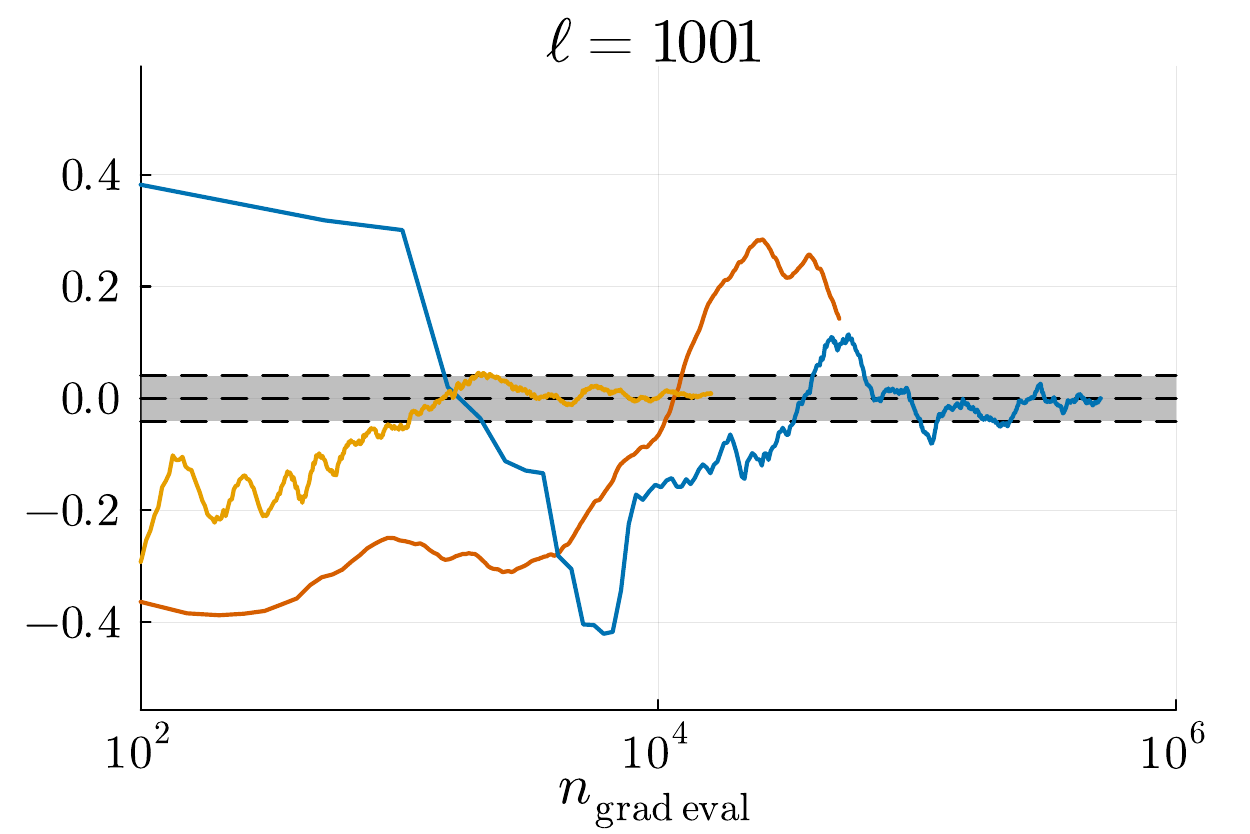}
    \end{subfigure}
    \caption{Relative bias of the cumulative mean of each sampler with respect to the NUTS reference values, normalized by the target standard deviation, as a function of the number of gradient evaluations. The three panels show representative multipoles at \(\ell=10,95,1001\). The curves are computed separately for each chain and then averaged to display the overall trend. The gray shaded region corresponds to the standard Monte Carlo error of NUTS, used as a benchmark. On average, the final bias introduced by MCLMC remains below this reference level, making it negligible and confirming the reliability of the sampler.}
    \label{fig:mean_bias_gradevals}
\end{figure}

\subsection{Samplers' performance}
\label{sec:sampler_perf}

\begin{figure}[!htp]
\centering
\includegraphics[width=0.75\textwidth]{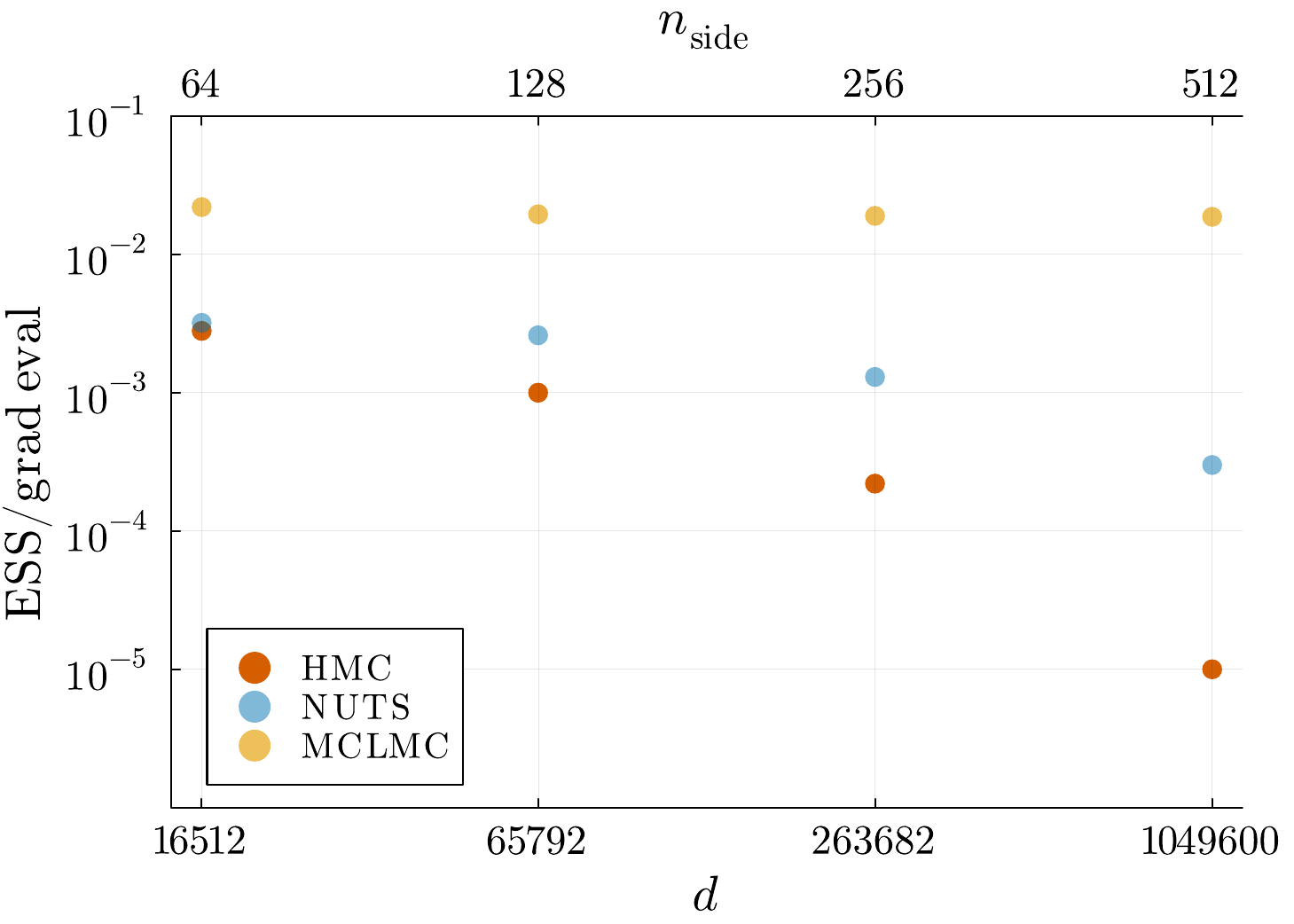}
\caption{Sampler performance versus problem size. We report efficiency as ESS per gradient evaluation as a function of dimensionality, for HMC, NUTS, and MCLMC. Performance degrades sharply with dimension for NUTS and especially HMC, whereas MCLMC exhibits only a mild decline, indicating markedly better scaling in high dimensions.}
\label{fig:ess_grad}
\end{figure}
Since the field-level approach is known to be an extremely promising yet computationally demanding technique, due to the high dimensionality of the parameter space, improving its performance is of vital importance if it is to plausibly replace traditional methods~\citep{Seljak:2017rmr}. For this reason, we explored samplers capable of delivering superior performance and making the entire pipeline more efficient. For example, Pathfinder can be used to initialize chains and, in future developments, to estimate and initialize the mass matrix for HMC and NUTS, thereby greatly reducing not only the burn-in phase, but also the number of tuning steps. More importantly, we tested a new family of MCMC methods, the MicroCanonical ones. MCLMC has proven to be a valuable, accurate, and high-performing tool for this class of problem. Since the computational cost of a gradient-based sampler is generally tied to the number of gradient evaluations it must perform, we assessed sampler performance using the average effective sample size per gradient evaluation (ESS/grad eval). Fig.~\ref{fig:ess_grad} shows the behaviour of this metric as a function of the problem dimensionality (or, equivalently, the map resolution). The results speak for themselves: MCLMC significantly outperforms the other two samplers, which not only have a lower ESS per gradient evaluation at fixed dimensionality, but also exhibit a much steeper decreasing trend, with HMC trailing behind. In the tested configuration with \(n_\mathrm{side}=512\), MCLMC performs almost 100 times better than NUTS and nearly 1,000 times better than standard HMC. 
To also provide an idea of the computational time cost, we report the average wall time required to sample a single chain for the three samplers at the configuration \(n_\mathrm{side}=512\). On average, one of our HMC chains takes two days (\(1.6\cdot10^5\) seconds) to run, one of NUTS takes six days (\(5\cdot10^5\) seconds), and one of MCLMC takes 7 hours (\(2.6\cdot10^4\) seconds). However, since we sampled chains of different lengths, for a fair comparison we also report the number of ESS per second: HMC is \(0.0003\), NUTS is \(0.0008\), and MCLMC is \(0.05\).
Therefore, when MCLMC is combined with a fully automatically differentiable model, the result is an extremely efficient and flexible pipeline. This does not change the fact that, given the current mathematical understanding of this sampler, careful validation remains necessary to assess any potential biases arising from the lack of asymptotic guarantees in MCLMC.

\end{document}